\documentclass[twocolumn]{jpsj3}
\usepackage[dvipdfm]{color}
\usepackage{txfonts}
\usepackage{bm}

\title{Anisotropic spin fluctuations in the quasi one-dimensional frustrated magnet $\mathrm{LiCuVO_4}$}

\author{Kazuhiro NAWA$^{1,2}$\thanks{E-mail address: knawa@issp.u-tokyo.ac.jp}, Masashi TAKIGAWA$^{2}$
\thanks{E-mail address: masashi@issp.u-tokyo.ac.jp}, \\
Makoto YOSHIDA $^{2}$, and  Kazuyoshi YOSHIMURA$^{1}$
\thanks{E-mail address: kyhv@kuchem.kyoto-u.ac.jp}, %\\
}

\inst{
$^{1}$Department of Chemistry, Graduate School of Science, Kyoto University, Kyoto 606-8502, Japan \\
$^{2}$Institute of Solid State Physics, The University of Tokyo, Kashiwanoha, Kashiwa, Chiba 277-8581, Japan
}

\abst{
We report results of NMR experiments on a single crystal of the quasi one-dimensional frustrated magnet 
$\mathrm{LiCuVO_4}$. The NMR spectra of $^7$Li and $^{51}$V nuclei indicate a helical spin order in a 
magnetic field of 4~T with the helical spin plane perpendicular to the field and a spin-density-wave 
(SDW) order at 10~T with modulation in the magnitude of the moments aligned along the field, 
in agreement with earlier reports. The nuclear spin-lattice relaxation rate $1/T_1$ at $^{51}$V nuclei, 
which is selectively coupled to the transverse spin fluctuations perpendicular to the field, 
shows a pronounced peak near the helical ordering temperature in the field of 4~T applied along the $a$-axis.  
In the field of 10~T, however, such a peak is absent. Instead $1/T_1$ at $^7$Li nuclei probing longitudinal spin 
fluctuations shows divergent behavior towards the SDW ordering temperature. These results are qualitatively 
consistent with the theoretical description that the SDW correlation is due to bound magnon pairs, which produce 
an energy gap in the transverse spin excitation spectrum.  
}

\kword{$\mathrm{LiCuVO_4}$, spin fluctuations, a frustrated magnetic chain, spin-lattice relaxation rate}

\begin{document}
\maketitle

\section{Introduction}

Frustrating interactions and quantum fluctuations in low dimensional spin systems prevent conventional 
magnetic order and can lead to exotic ground states such as spin-liquids\cite{SL}, valence bond 
crystals\cite{SL, VBC}, spin-nematic, or more generally, spin-multipolar ordered states\cite{nematic, nematic2,
octupole, 1Dnematic0, 1Dnematic, 1Dnematic2}. In magnetic fields, where the Zeeman interaction tends 
to align the spins and competes with quantum fluctuations and frustration, novel phases 
such as magnetization plateaus\cite{plateau, SCBONMR, plateau2} are expected in the magnetization process. 
  
In this respect, one-dimensional (1D) spin 1/2 systems with the ferromagnetic nearest neighbor interaction 
$J_1$ frustrating with the antiferromagnetic next-nearest neighbor interaction $J_2$ in a magnetic field $h$
\begin{equation}
\label{J1J2}
\mathcal{H} = J_1 \sum _{l} \mathbf{s}_l \cdot \mathbf{s}_{l+1} +
J_2 \sum _{l} \mathbf{s}_l \cdot \mathbf{s}_{l+2} - h \sum _{l} s_{l}^{z}
\end{equation}
provide a particularly interesting example,which has been extensively studied in recent theories~\cite{1Dnematic0, 1Dnematic, 1Dnematic2, 1Dtheory0,1Dtheory1,1Dtheory2,1Dtheory3,1DtheoryofT11,1DtheoryofT12}.  
The ground state of this model for classical spin chains has a helical magnetic order at zero magnetic field, 
which is destabilized by quantum fluctuations for spin 1/2. In finite fields, however, helical spin correlation 
in quantum systems leads to a long-range order of the vector-chirality defined as  
$\mathbf{\kappa_{l}}^{(n)} = \mathbf{s}_l \times \mathbf{s}_{l+n}$, ($n$ = 1 or 2). 
As the field increases, dominant spin correlation changes due to formation of bound states of 
two or more magnons. This leads to a quasi-long-range order of a spin-density-wave (SDW) correlation 
$\langle s_i^z s_j^z \rangle - \langle s_i^z \rangle \langle s_j^z \rangle$ along the 
field ($z$) direction or a bond nematic correlation 
$\langle s_i^+ s_{i+1}^+ s_j^- s_{j+1}^- \rangle$ 
% (more generally the $p$-th order multipolar correlations $\langle s_i^+ s_{i+1}^+ ...  s_{i+p-1}^+s_j^- s_{j+1}^- ...  s_{i+p-1}^- \rangle$) 
perpendicular to the field. 

\begin{figure}[b]
\centering
\includegraphics[height=5cm]{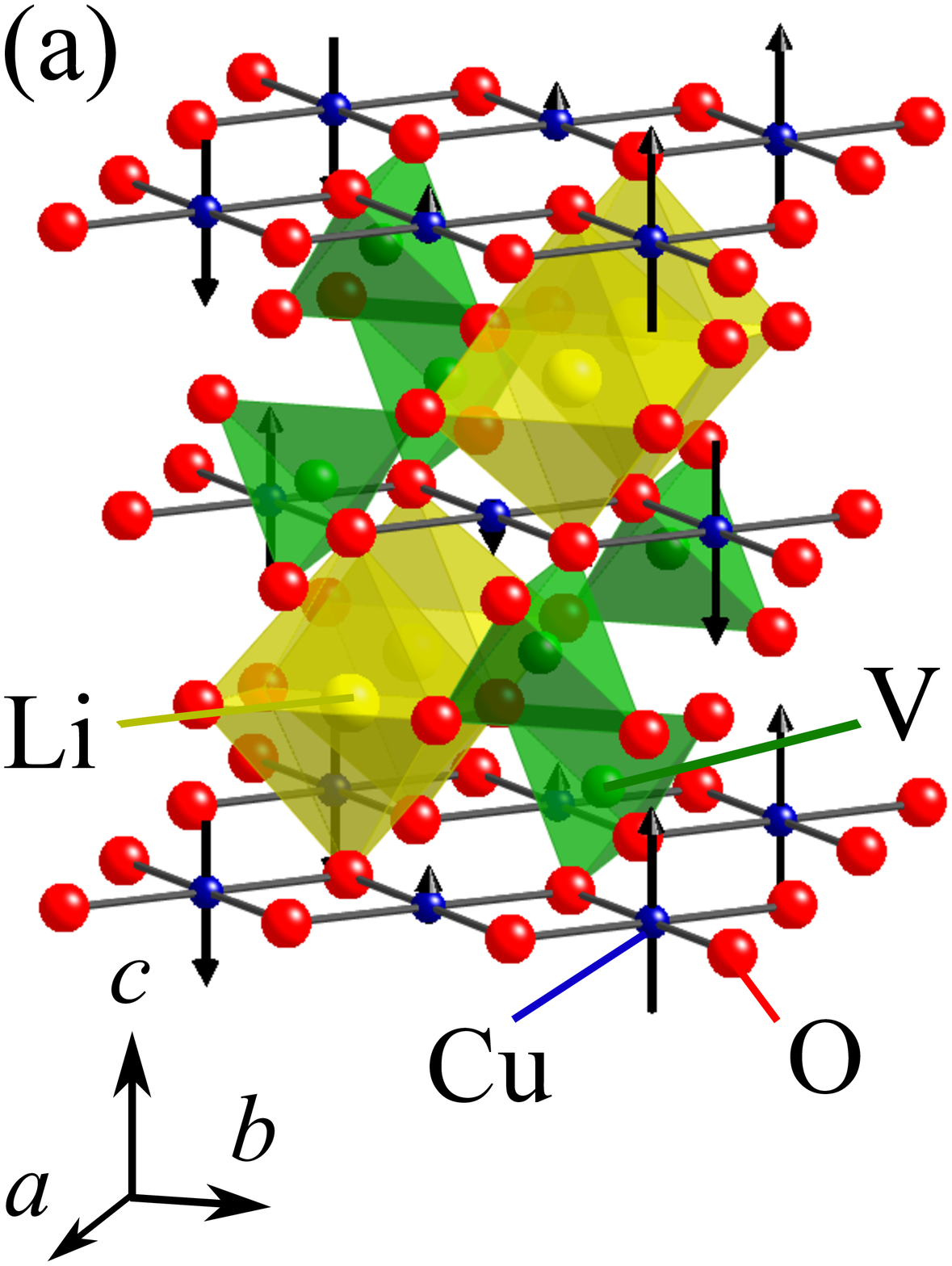} 
\includegraphics[height=5cm]{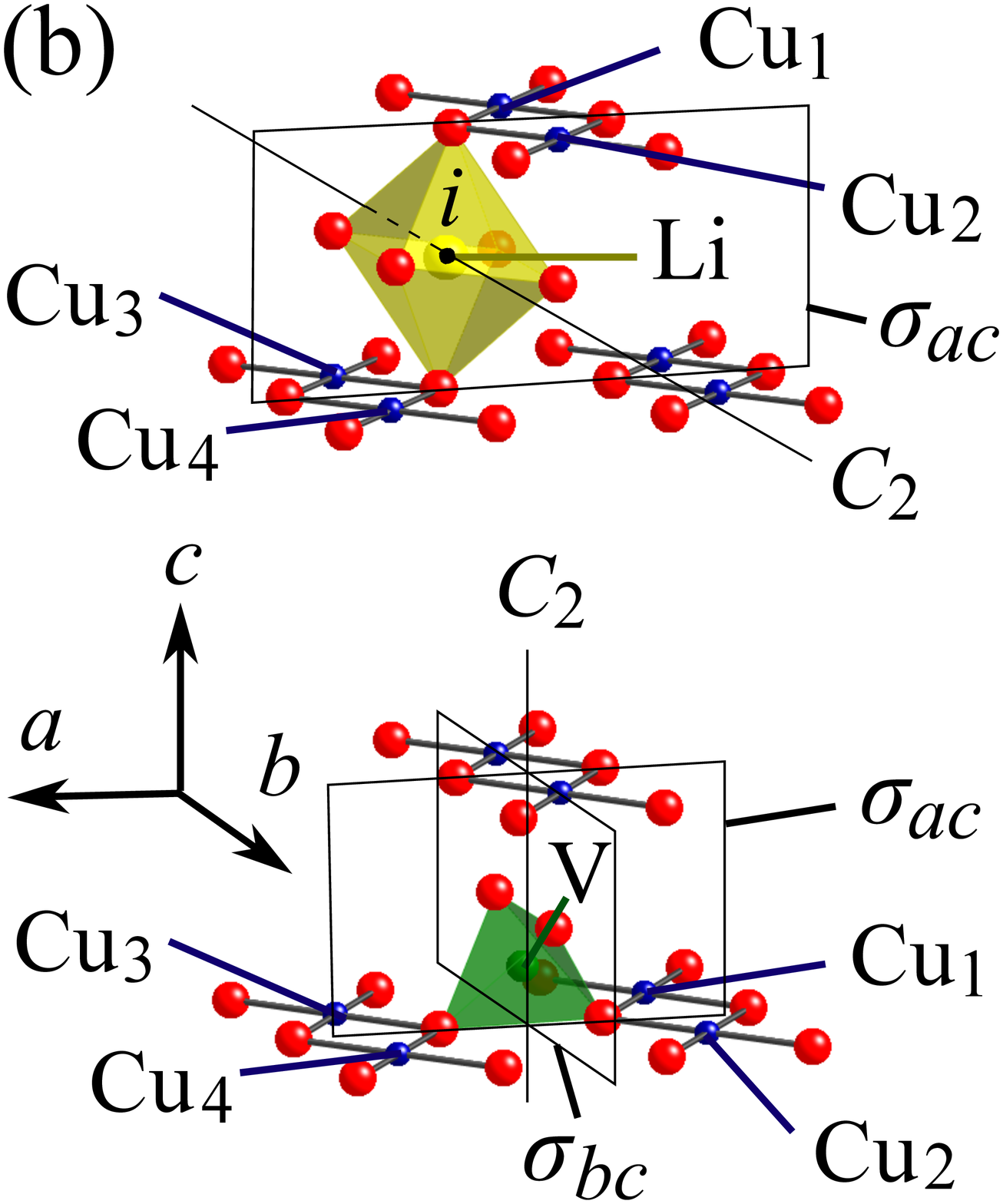} 
\caption{ (Color Online) 
(a) The crystal structure of $\mathrm{LiCuVO_4}$. The Cu and O ions form frustrated chains along the $b$-axis.
The Li and V ions occupy the center of octahedra (yellow) and tetrahedra (green), respectively.
The arrows illustrate the spin structure in the SDW phase for $H \parallel c$.
(b) Local site symmetry of the Li and V sites. For both sites, four nearest neighbor Cu ions are 
related each other by the symmetry operations (twofold rotation $C_2$ or mirror operations 
$\sigma_{\mu \nu}$), that leave the nuclear site unchanged.}
\label{crystal}
\end{figure}

The quasi-long-range order in pure 1D systems can turn into a true long range order in real materials with 
interchain interactions. A large number of cuprate materials containing frustrated $J_1$-$J_2$ chains have been 
known to date, of which the best studied example is $\mathrm{LiCuVO_4}$\cite{sample1, sample2, neutron0,
inelastic, neutron1, neutron2, neutron3, NMR2, NMR3, NMR4, magnetization, cryst, ESR}. 
It has an orthorhombic crystal structure with the space group $Imma$\cite{cryst} as shown in 
Fig.~\ref{crystal}(a) and contains chains of Cu$^{2+}$ (spin 1/2) formed by edge-sharing $\mathrm{CuO_4}$ 
plaquettes. In this geometry, the nearest neighbor Cu ions are coupled by two Cu-O-Cu bonds with approximately
$90^\circ$ bond angle, leading to a ferromagnetic interaction. In addition, a sizable antiferromagnetic
superexchange is expected between the next-nearest neighbor Cu ions through two Cu-O-O-Cu paths.
There are two such chains in an unit cell extending along the $b$-axis separated by $(\mathbf{a}+\mathbf{c})/2$.

At zero field, $\mathrm{LiCuVO_4}$ shows an incommensurate helical magnetic order below $T_N$ = 2.1~K
with the magnetic scattering vector $\mathbf{Q}_0 = 2 \pi (1, 0.468, 0)$ and magnetic moments lying 
in the $ab$-plane\cite{neutron0}. Previous NMR\cite{NMR2, NMR3, NMR4} and neutron diffraction 
experiments\cite{neutron1, neutron2, neutron3} revealed that the spin structure changes with magnetic field
$H$. A spin flop transition occurs at $H$ = 2.5~T, flipping the moments into the plane perpendicular to the
field. Above 7~T a longitudinal SDW order is observed, where the moments are aligned parallel to the
field with their magnitudes modulated along the chains. Furthermore, a magnetization curve shows 
an abrupt change of slope at 41~T (for $H \parallel c$) slightly below the saturation field of 
45~T\cite{magnetization}. This may be a signature for a nematic order that breaks the spin rotational 
symmetry by spontaneous development of anisotropic spin correlation but preserves the time reversal symmetry. 
The exchange parameters were proposed as $J_1 = -18.5$ K and $J_2 = 44$ K based on the analysis of the spin 
wave dispersion at zero field\cite{inelastic}. However, they disagree strongly with the values 
$J_1 = -182$ K and $J_2 = 91$ K estimated by fitting the magnetic susceptibility data\cite{susceptibility}
and controversy still remains.   

Theories have proposed that the key element behind the SDW correlation in the model of Eq.~\eqref{J1J2} 
as well as in $\mathrm{LiCuVO_4}$ is the bound state of magnon pairs stabilized by ferromagnetic $J_1$~\cite{1Dtheory0, 1Dtheory2, 1Dtheory3}. 
This has direct consequences on the magnetization process, i.e. the magnetization changes by a step of 
$\Delta s_z = 2$, and on the relation between the SDW wave vector and the magnetization as was confirmed by
neutron diffraction experiments on $\mathrm{LiCuVO_4}$~\cite{neutron2, neutron3}. Furthermore, the low energy
spin dynamics is fundamentally changed. At low fields, incommensurate transverse spin fluctuations are the
dominant low energy excitations, leading to the helical order. At higher fields, however, an energy gap due
to bound magnon pairs develops in the transverse spin excitations and makes the longitudinal fluctuations
dominant, leading to the SDW order.

Direct observation of spin dynamics is thus important for microscopic understanding of the field-induced 
phase transitions in quasi 1D frustrated magnets. To our knowledge, however, systematic measurements of 
spin dynamics as a function of magnetic field has not been reported yet. The nuclear spin-lattice 
relaxation rate 1/$T_1$ measured by nuclear magnetic resonance (NMR) experiments is a particularly 
powerful probe for such a purpose. Indeed, theories have made specific predictions on temperature and 
field dependences of 1/$T_1$ for the model in Eq.~ \eqref{J1J2}~\cite{1DtheoryofT11,1DtheoryofT12}. 
In this paper, we report results of $^{7}$Li and 
$^{51}$V NMR experiments on a single crystal of $\mathrm{LiCuVO_4}$.      

The paper is organized as follows. Experimental details on sample synthesis and NMR measurements are 
described in Section 2. The $^{7}$Li and $^{51}$V NMR spectra in the paramagnetic, helical and SDW phases 
are presented in Section 3. The NMR spectra agree with the earlier results by  B\"{u}ttgen \textit{et al.}\cite{NMR2,NMR3}.
Here we describe how one can deduce the spin structure from NMR spectra based on symmetry 
properties of hyperfine coupling tensors. 
In Section 4, we discuss the anisotropic spin fluctuations based on
temperature and field dependences of $1/T_1$. First we present analysis of hyperfine form factors, which 
indicate that transverse spin fluctuations can be effectively probed by $^{51}$V nuclei, while longitudinal 
fluctuations are better probed by $^{7}$Li nuclei. At $H$ = 4~T, where the ground state 
has a helical order, the transverse fluctuations are dominant, making $1/T_1$ at V sites diverge towards    
$T_N$. On the other hand, at $H$ = 10~T, where the ground state has a SDW order, contribution from the 
transverse fluctuations to $1/T_1$ at V sites decreases with decreasing temperature without any anomaly at 
$T_N$. Instead $1/T_1$ at Li sites shows a pronounced peak at $1/T_1$ due to the longitudinal fluctuations.
These results are qualitatively consistent with the theoretical predictions\cite{1DtheoryofT12}.
Conclusions are presented in Section 5.  

\section{Experiments}
Single crystals of $\mathrm{LiCuVO_4}$ were grown by a flux method using $\mathrm{LiCuVO_4}$, 
$\mathrm{LiVO_3}$ and $\mathrm{LiCl}$ as starting materials\cite{sample1, sample2}. 
Powder samples of $\mathrm{LiCuVO_4}$ and $\mathrm{LiVO_3}$ were synthesized by solid-state reaction: $\mathrm{Li_2CO_3}$ (5N), $\mathrm{V_2O_5}$ (4N) and $\mathrm{CuO}$ (5N) were mixed in 
a stoichiometric ratio, pelletized and then sintered at 550 $^\circ$C for 96 -144 hours with an 
intermediate grinding. For growth of single crystals, 
$\mathrm{LiCuVO_4}$, $\mathrm{LiVO_3}$ and $\mathrm{LiCl}$ (4N) were mixed in a molar ratio of 0.25 : 0.40 : 0.35. The mixture was melt at 600 $^\circ$C and then slowly cooled down to 500 $^\circ$C in 100 hours.
Single crystals were separated from the solidified melt by washing the melt in hot water.
Crashed powders of single crystals were confirmed to be in a single phase by powder X-ray diffraction measurements. Crystal axes were determined by using an imaging plate diffractometer in a oscillation mode.
The magnetic susceptibility was measured by a SQUID magnetometer in a magnetic field of 4 T.

NMR measurements for $^{7}$Li and $^{51}$V nuclei were performed on a single crystal with the size 1.0 
$\times$ 1.2 $\times$ 0.5 $\mathrm{mm}^3$ using the spin echo technique. The Li atoms occupy the
$4d$ sites located at the center of the octahedra connecting two $\mathrm{CuO_2}$ chains displaced by 
$\left(\mathbf{a}+\mathbf{c}\right)/2$. The V atoms occupy the $4e$ sites at the 
center of the tetrahedra connecting two chains displaced by $\mathbf{a}$. The local symmetries of these 
sites are also shown in Fig.~\ref{crystal}(b). In the paramagnetic state above 6 K, NMR spectra of both
$^{7}$Li and $^{51}$V nuclei were obtained by Fourier transforming the spin echo signals at a single 
rf-frequency. The fundamental parameters of $^{7}$Li and $^{51}$V nuclei, the nuclear spin $I$, the 
gyromagnetic ratio $\gamma_N$, and the nuclear quadrupole moment $Q$, are listed in Table~\ref{nuQ}.  

\begin{figure}
\centering
\includegraphics[height=6.9cm,clip]{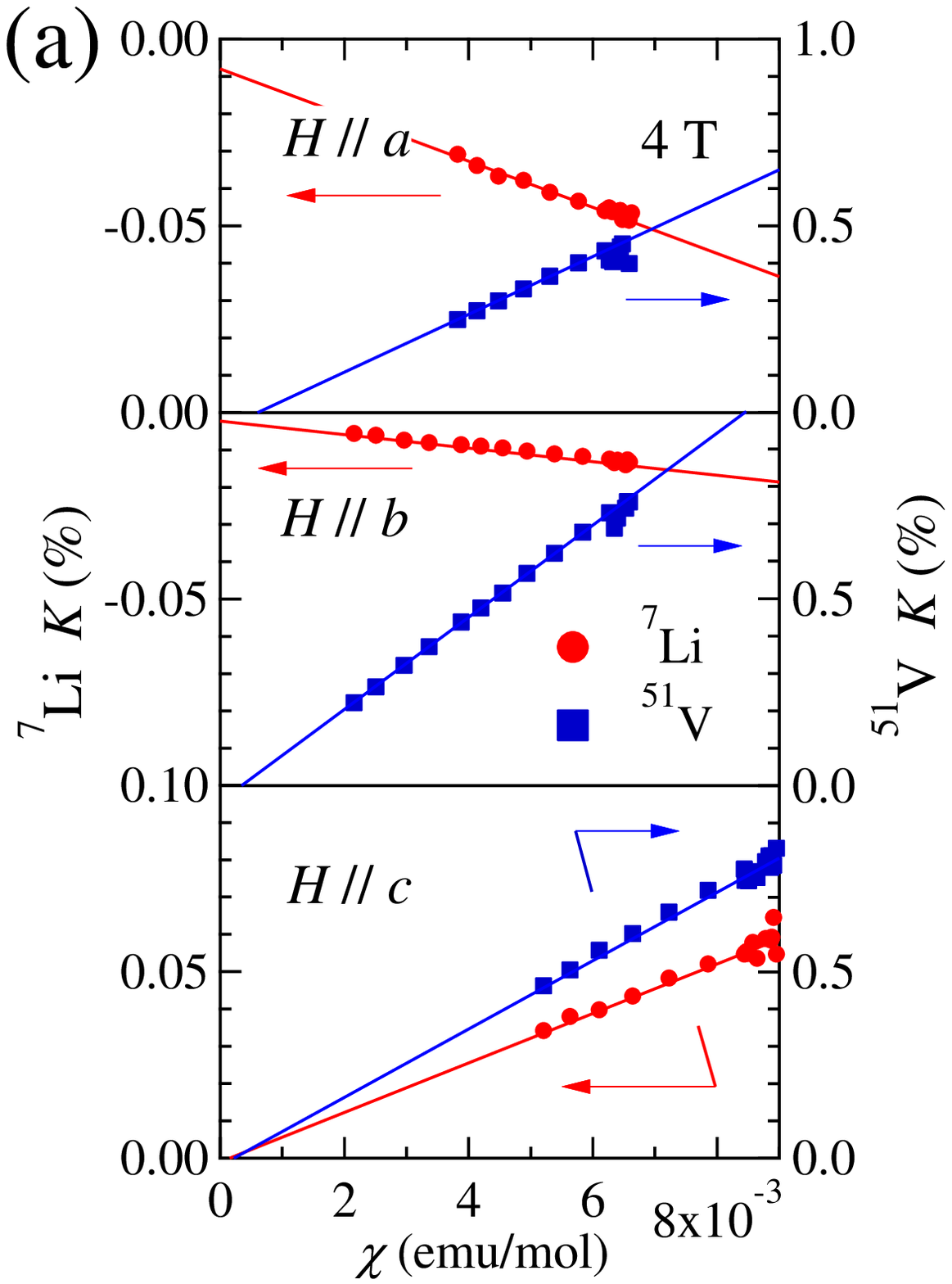} 
\includegraphics[height=6.9cm,clip]{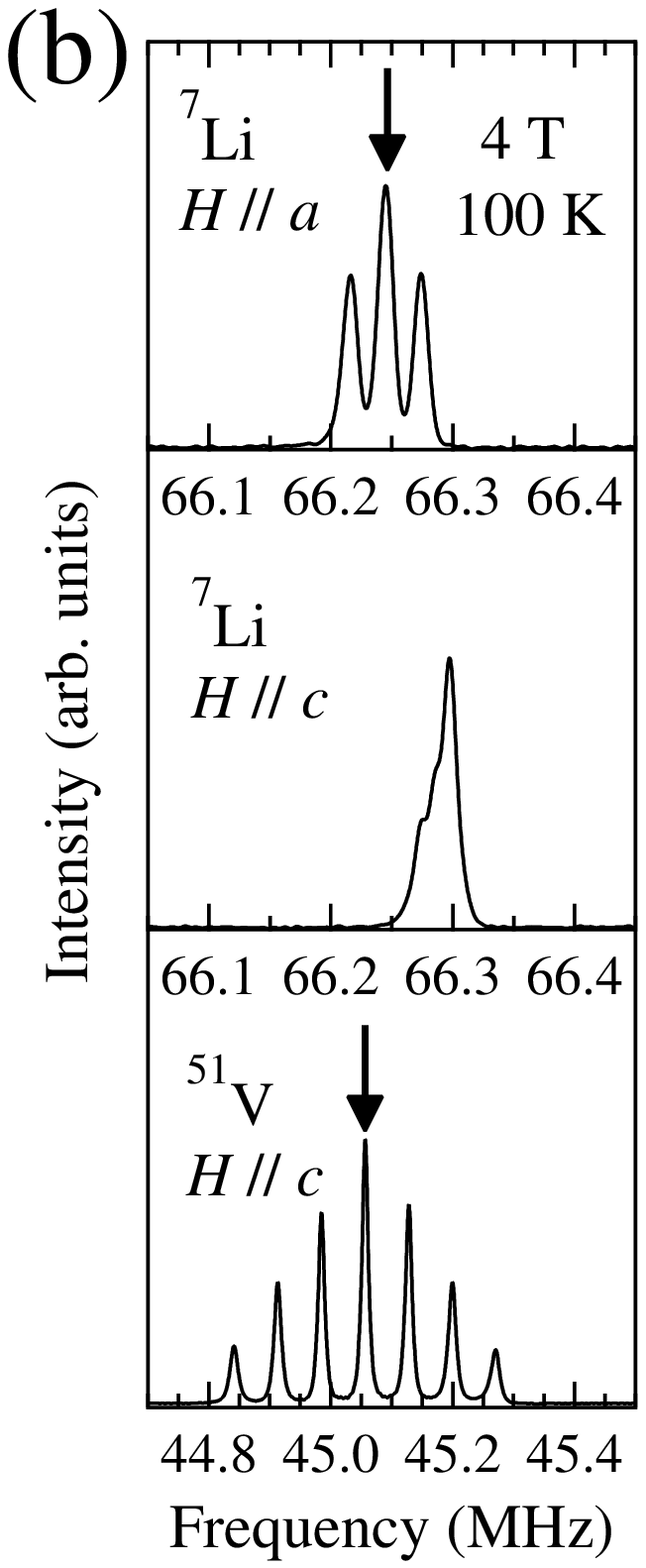}
\caption{ (Color Online) 
(a) The $K$-$\chi$ plots for $^7$Li (red circles) and $^{51}$V (blue squares) nuclei. The diagonal components 
of the hyperfine coupling tensors determined by fitting the $K$-$\chi$ plots to Eq. \eqref{Kchi2}
are listed in Table~\ref{hyperfine}. (b) Typical $^7$Li and $^{51}$V NMR spectra in the paramagnetic state.
The magnetic shift $K$ is determined from the central peak of the quadrupole split spectra indicated by the arrows, except that the shift of $^7$Li nuclei for $H \parallel c$ is determined from the center of gravity of the whole spectra.}
\label{Kchiplot4T}
\end{figure}

Typical NMR spectra in the paramagnetic state are shown in Fig.~\ref{Kchiplot4T}(b).
When the magnetic field is applied along the crystalline $a$-, $b$- or $c$-axis, all Li or V atoms 
become equivalent. The NMR spectra then consist of seven peaks for $^{51}$V and three peaks for $^7$Li, 
which are split by the nuclear quadrupole interaction. The quadrupole splitting 
$\nu_{\alpha \alpha}$ ($\alpha=a, b,$ or $c$), that is the interval between neighboring peaks, is related to 
the diagonal components of the electric field gradient (EFG) tensor $V_{\alpha \alpha}$ as 
\begin{equation}
\nu_{\alpha \alpha} = \frac{3 e Q V_{\alpha \alpha}}{2I (2I - 1) h},
\end{equation}
where $h$ is the Planck constant. The values of $\nu_{\alpha \alpha}$ are listed in Table~\ref{nuQ}.
For the V sites, the crystalline $a$-, $b$- and $c$-axes are the principal axes of the EFG tensor, 
therefore, the quadrupole splitting takes extremal (minimum or maximum) values along these direction.
This situation allows us to precisely orient the crystal in the magnetic field by using a double-axis 
goniometer.
For the Li sites, while the $b$-axis is one of the principal axes of 
EFG, the directions of two other principal axes in the $ac$-plane remain unknown.
Magnetic shifts of $^{7}$Li and $^{51}$V
nuclei were determined from the peak frequency of the central line except for $^{7}$Li with 
$H \parallel c$, where the magnetic shift was determined from the center of gravity 
of the whole spectra since the quadrupole splitting is too small to be resolved clearly.

Below 6~K, the NMR spectra of both $^{7}$Li and $^{51}$V nuclei become so broad that 
rf-pulses at a single frequency do not provide sufficient band width.  
The $^{51}$V spectra were then obtained by sweeping the magnetic field with a fixed 
rf-frequency and the spin-echo intensity was recorded as a function of the magnetic field.
The $^{7}$Li spectra, on the other hand, were obtained by summing the Fourier transform of the spin echo 
signal obtained at different frequencies with a fixed magnetic field. 

The inversion-recovery method was used for the measurement of $1/T_1$. 
We determined $1/T_1$ by fitting the spin-echo intensity $M(t)$ as a function of
the time $t$ after the inversion pulse to a single or stretched exponential recovery function 
\begin{equation}\label{eq:stretch}
M(t) = M_{\rm eq}-M_0 \exp \left\{-(t/T_1)^{\beta} \right\},
\end{equation}
where $\beta$ is the stretch exponent that provides a measure of inhomogeneous distribution of 1/$T_1$.
When $\beta$ = 1 Eq.~\eqref{eq:stretch} reduces to a single exponential function corresponding to 
homogeneous relaxation. Near or below $T_N$, $1/T_1$ was measured at the center of the broad NMR spectra. 

\begin{table}
\caption{Fundamental properties and the quadrupole splitting for $^7$Li and $^{51}$V nuclei. 
Underlines indicate that they correspond to the principal values of the electric field gradient tensor.}
\label{nuQ}
\begin{center}
\begin{tabular}{lcccccc}
\hline
nuclei & $I$ & $\gamma/(2\pi)$ & Q & $\nu_{aa}$ & $\nu_{bb}$ & $\nu_{cc}$ \\
& & (MHz/T) & ($10^{-24}$ cm$^2$) & (MHz) & (MHz) & (MHz) \\
\hline
$^7$Li & 3/2 & 16.5468 & -4.0 & 0.0290 & \underline{0.0219} & $<$ 0.01 \\
$^{51}$V & 7/2 & 11.1988 & -5.2 & \underline{0.0160} & \underline{0.0879} & \underline{0.0715} \\
\hline
\end{tabular}
\end{center}
\end{table}

\section{NMR spectra}
\subsection{Paramagnetic state}
The magnetic shifts $K$ for $^7$Li and $^{51}$V nuclei in the paramagnetic state were measured 
as a function of temperature in the magnetic field of 4 T. The magnetic shift $K$ is defined as the 
difference between the local magnetic field $H_\mathrm{loc}$ acting on a nucleus 
and the external magnetic field $H_\mathrm{ext}$ normalized by $H_\mathrm{ext}$,
\begin{equation}
K(T) = \frac{H_\mathrm{loc} - H_\mathrm{ext}}{H_\mathrm{ext}}. \label{K}
\end{equation}
The value of $K$ is experimentally determined by substituting $H_\mathrm{loc} = \omega/\gamma_N$ 
for Eq.~\eqref{K}, where $\omega$ is a resonance frequency and $\gamma_N$ is a nuclear gyromagnetic ratio.

The local magnetic field $\mathbf{H}_\mathrm{loc}$ is given by the sum of the external field $\mathbf{H}_\mathrm{ext}$, the Lorentz field $\mathbf{H}_\mathrm{Loz}$, the demagnetization field
$\mathbf{H}_\mathrm{dem}$, and the hyperfine field $\mathbf{H}_\mathrm{hf}$ produced by electronic moments as 
\begin{equation}
	\begin{split}
\mathbf{H}_\mathrm{loc} &= \mathbf{H}_\mathrm{ext} + \mathbf{H}_\mathrm{Loz} + \mathbf{H}_\mathrm{dem} + \mathbf{H}_\mathrm{hf} \\
\mathbf{H}_\mathrm{Loz} &= \frac{4}{3} \pi \frac{\mathbf{M}}{N_\mathrm{A} v} \\
\mathbf{H}_\mathrm{dem} &= -4 \pi \mathbf{N} \cdot  \frac{\mathbf{M}}{N_\mathrm{A} v} \\
\mathbf{H}_\mathrm{hf} &= \frac{1}{N_\mathrm{A} \mu_B}(\mathbf{C}^\mathrm{tr} + \mathbf{C}^\mathrm{dip}) \cdot \mathbf{M}, \label{localfield}
	\end{split}
\end{equation}
where $\mathbf{N}$ is a demagnetization tensor, $\mathbf{M}$ is magnetization per mole,
$N_\mathrm{A}$ is the Avogadro number, $\mu_B$ is the Bohr magnetron and $v$ is volume per formula unit.
$\mathbf{C}^\mathrm{tr}$ and $\mathbf{C}^\mathrm{dip}$ are the hyperfine coupling tensors
due to transferred hyperfine and dipolar interactions, respectively.
In the paramagnetic state, $\mathbf{M} = \bm{\chi} \cdot \mathbf{H}_\mathrm{ext}$, 
where $\bm{\chi}$ is the susceptibility tensor. Therefore, 
\begin{equation}
\mathbf{H}_\mathrm{loc} - \mathbf{H}_\mathrm{ext} = \mathbf{K} \cdot \mathbf{H}_\mathrm{ext} ,
\label{Ktensor}
\end{equation}
where the shift tensor $\mathbf{K}$ is given by  
\begin{equation}
\mathbf{K} = \frac{1}{N_\mathrm{A} \mu_B} \left\{ \frac{4}{3} \pi \frac{\mu_B}{v} (\mathbf{1} - 3 \mathbf{N}) 
+ \mathbf{C}^\mathrm{tr} + \mathbf{C}^\mathrm{dip} \right\} \cdot \bm{\chi}.
\label{Kchi}
\end{equation}

The value of the shift defined in Eq.~\eqref{K} under the field along the $i$-axis ($i$ = $a$, $b$, $c$) is  
given by the diagonal component $K_{ii}$ of the shift tensor in Eq.~\eqref{Ktensor}. Since $\bm{\chi}$ is 
diagonal in the coordinate system of crystalline axes, we obtain
\begin{equation}
	\begin{split}
K_{ii}(T) &= \frac{C_{ii}}{N_\mathrm{A} \mu_B} \chi_{ii}(T) \\
C_{ii} &= \frac{4}{3} \pi \frac{\mu_B}{v} (1 - 3 N_{ii}) + C^\mathrm{tr}_{ii} + C^\mathrm{dip}_{ii},
\label{Kchi2}
	\end{split}
\end{equation}
where $N_{ii}$, $C^\mathrm{tr}_{ii}$, $C^\mathrm{dip}_{ii}$ are the diagonal components 
of $\mathbf{N}$, $\mathbf{C}^\mathrm{tr}$ and $\mathbf{C}^\mathrm{dip}$, respectively.
Here, $C_{ii}$ can be experimentally determined by measuring
both the magnetic shift $K_{ii}$ and the magnetic susceptibility $\chi_{ii}$.

Figure~\ref{Kchiplot4T}(a) shows the $K-\chi$ plots for $^7$Li and $^{51}$V nuclei with linear fits
indicated by the solid lines. The slope determines the values of $C_{ii}$ and the intersection is related to the temperature-independent van Vleck (orbital) contribution to $K$ and $\chi$.  
The values of $C_{ii}$ are listed in Table \ref{hyperfine}. We also show the values of $C^\mathrm{dip}_{ii}$ 
obtained by the lattice-sum calculation. While $C_{ii}$ for $^{51}$V is an order of magnitude larger than
$C^\mathrm{dip}_{ii}$, indicating that the dominant contribution comes from the transferred hyperfine 
couping $C^\mathrm{tr}_{ii}$, this is not the case for $^7$Li. In order to determine $C^\mathrm{tr}_{ii}$, 
we subtract contributions from the demagnetization and dipolar fields from the experimental values of $C_{ii}$. 
The diagonal components of the demagnetization 
tensor $N_{ii}$ are estimated by approximating the sample shape with a spheroid\cite{demagnetization}. 
In Table \ref{hyperfine}, we also show 
\begin{equation}
C^\mathrm{tr}_{ii} = C_{ii} - \frac{4}{3} \pi \frac{\mu_B}{v} (1 - 3 N_{ii}) - C^\mathrm{dip}_{ii}. 
\label{Ctr}
\end{equation}
The $C^\mathrm{tr}_{ii}$ for $^{7}$Li nuclei is much smaller than $C^\mathrm{dip}_{ii}$ and should be neglected,
considering the errors involved in the estimation of $N_{ii}$. Therefore, we conclude that 
only dipolar interactions contribute to the coupling tensor of $^7$Li nuclei.
This is reasonable since light elements such as Li usually have small covalency with magnetic ions. 
For $^{51}$V nuclei, $C^\mathrm{tr}_{ii}$ is much larger than $C^\mathrm{dip}_{ii}$ for all directions,
which ensures that the dominant source of hyperfine field is the short ranged transferred hyperfine 
interactions. 

\begin{table}[b]
\caption{$C_{ii}$: the diagonal components of the hyperfine coupling tensors obtained from the $K$-$\chi$ plots
and Eq.~\eqref{Kchi2}. $C_{ii}^\mathrm{dip}$: the dipolar coupling calculated by lattice sum within a  
sphere with the radius of 60 \AA. $C_{ii}^\mathrm{tr}$: the contribution from the transferred hyperfine 
interactions estimated from Eq.~\eqref{Ctr}. $A_{ii}^{(1)}$: The hyperfine coupling to a nearest 
neighbors spin including both the transferred hyperfine and the dipolar contributions.}
\label{hyperfine}
\begin{center}
\begin{tabular}{lcccc}
$^7$Li & & & & \\
\hline
$i$ & $C_{ii}$ (T/$\mu_B$) & $C_{ii}^\mathrm{dip}$ (T/$\mu_B$) & $C_{ii}^\mathrm{tr}$ (T/$\mu_B$) & \\
\hline
$a$ & -0.035(1) & -0.035 & 0.008 & \\
$b$ & -0.009(0) & -0.032 & -0.002 & \\
$c$ & 0.037(1) & 0.067 & 0.003 & \\
\hline \\
$^{51}$V & & & & \\
\hline
$i$ & $C_{ii}$ (T/$\mu_B$) & $C_{ii}^\mathrm{dip}$ (T/$\mu_B$) & $C_{ii}^\mathrm{tr}$ (T/$\mu_B$) & $A_{ii}^\mathrm{(1)}$ (T/$\mu_B$)\\
\hline
$a$ & 0.432(3) & 0.063 & 0.360 & 0.119 \\
$b$ & 0.688(3) & -0.043 & 0.707 & 0.166 \\
$c$ & 0.533(5) & -0.020 & 0.586 & 0.129 \\
\hline
\end{tabular}
\end{center}
\end{table}

\subsection{Helical state}
\begin{figure*}[t]
\centering
\includegraphics[width=17cm,clip]{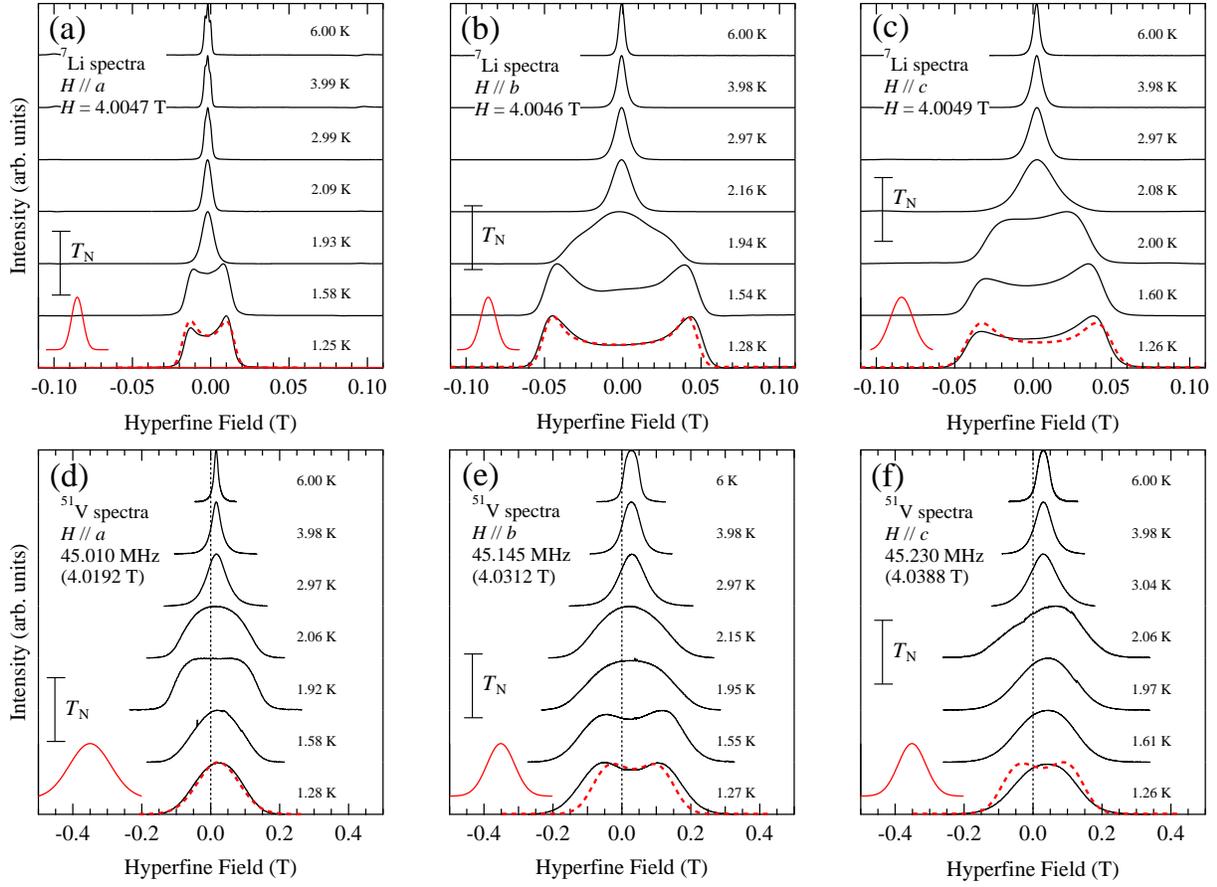}
\caption{ (Color Online) 
Frequency-swept NMR spectra of $^7$Li nuclei for (a) $H \parallel a$, (b) $H \parallel b$, (c) $H \parallel c$
and field-swept NMR spectra of $^{51}$V nuclei for (d) $H \parallel a$, (e) $H \parallel b$ and (f) $H \parallel c$ in the magnetic field near 4 T. The precise values of the fixed magnetic field are indicated in Fig. (a-c). 
The field values in the parenthesis in Fig. (d-f) correspond to the origin of the hyperfine field $\omega/\gamma_N$. The vertical bars represent uncertainty of $T_N$. The red dashed curves are simulated spectra for the helical phase where magnetic moments aligned perpendicular to the magnetic field. 
The red solid curves in the left side of the figures show
the Gaussian line shape used for convolution to simulated NMR spectra.}
\label{spectra4T}
\end{figure*}
NMR spectra of $^7$Li and $^{51}$V nuclei in the magnetic field of 4~T are shown as black solid curves
in Fig. \ref{spectra4T}(a-c) and (d-f) at various temperatures.
Here the frequency-swept $^7$Li spectra and the field-swept $^{51}$V spectra
are converted as a function of the hyperfine field by the following relation,
\begin{equation}
H_\mathrm{hf} \sim H_\mathrm{loc} - H_\mathrm{ext} = \frac{\omega}{\gamma_N} - H_\mathrm{ext}.
\end{equation}
At both sites, a sharp single NMR line at high temperatures changes to broad 
spectra at low temperatures, indicating a magnetic ordering transition near 2~K. 
The change of the NMR spectral shape is particularly pronounced at the $^7$Li sites. 
The spectra at low temperatures have a clear double-horn type line shape, which is characteristic of an 
incommensurate helical or SDW order. The transition temperatures are indicated by
the vertical bars in Fig.~\ref{spectra4T}(a-f) denoted $T_N$, at which the  $^7$Li line shape 
changes most rapidly. They agree with those determined in the previous NMR measurement\cite{NMR2}.

From the previous NMR and neutron diffraction experiments, a helical spin structure 
was concluded where magnetic moments lie perpendicular to the magnetic field\cite{NMR2, neutron1, neutron2}.
Magnitude of the magnetic moment and the ordering wave vector are determined by the 
neutron experiments as $\mathbf{\mu} = 0.31 \mu_B$ and $\mathbf{Q}_0 = 2 \pi (1, 0.468, 0)$\cite{neutron1, neutron2}.
In this section, we analyze details of the NMR spectra based on the local symmetry of the probing nuclei\cite{NMRsymmetry}. 

\subsubsection{$^7$Li nuclei}
We now discuss how one can understand line shapes of the NMR line shape qualitatively from the symmetry
of the hyperfine interaction. Figure \ref{crystal}(b) shows the configuration of Cu spins around one Li nucleus.
The hyperfine field $\mathbf{H}_\mathrm{hf}$ at a $^{7}$Li nucleus produced by
the surrounding electron spins $\mathbf{S}(\mathbf{r}_i)$ located at the Cu site $\mathbf{r}_i$ can be expressed as 
\begin{equation}
\mathbf{H}_\mathrm{hf} = \sum_i \mathbf{A}(\mathbf{r}_i) \cdot \mathbf{g} \cdot \mathbf{S}(\mathbf{r}_i), \label{Hint}
\end{equation}
where $\mathbf{A}(\mathbf{r}_i)$ is the hyperfine coupling tensor between the nucleus and $\mathbf{S}(\mathbf{r}_i)$.
The $\mathbf{g}$ is the $g$ tensor
\begin{equation}
\mathbf{g} = \begin{pmatrix}
g_{aa} & 0 & 0 \\
0 & g_{bb} & 0 \\
0 & 0 & g_{cc} \end{pmatrix}.
\end{equation}
Using the Fourier components of the coupling tensor and the electron spins
\begin{equation}
\mathbf{A}(\mathbf{q}) = \sum_\mathbf{i} \mathbf{A}(\mathbf{r}_i) e^{i\mathbf{q} \cdot \mathbf{r}_i}, \ \ 
\mathbf{S}(\mathbf{q}) = \frac{1}{\sqrt{N}} \sum_\mathbf{i} \mathbf{S}(\mathbf{r}_i) e^{-i\mathbf{q} \cdot \mathbf{r}_i}, \label{FT}
\end{equation}
the hyperfine field is rewritten as
\begin{equation}
\mathbf{H}_\mathrm{hf} = \frac{1}{\sqrt{N}} \sum_{\mathbf{q}} \mathbf{A}(\mathbf{q}) \cdot \mathbf{g} \cdot \mathbf{S}(\mathbf{q}), %\tag{A.3}
\label{A.3}
\end{equation}
where $N$ is the number of Cu spins in the system.

Let us first consider the coupling tensor $\mathbf{A}^{(1)}(\mathbf{r}_i) \ (i = 1 - 4)$ 
between the $^{7}$Li nucleus and the four nearest neighbor Cu spins shown in  Fig.~\ref{crystal}(b). 
We write $\mathbf{A}^{(1)}(\mathbf{r}_1)$, the coupling to the spin on Cu1, as 
\begin{equation}
\mathbf{A}^{(1)}(\mathbf{r}_1) = \begin{pmatrix}
A_{aa}^{(1)} & A_{ab}^{(1)} & A_{ac}^{(1)} \\
A_{ba}^{(1)} & A_{bb}^{(1)} & A_{bc}^{(1)} \\
A_{ca}^{(1)} & A_{cb}^{(1)} & A_{cc}^{(1)} \end{pmatrix}.% \ \ 
%\mathbf{S}_i = \begin{pmatrix}
%S_{a} \\
%S_{b} \\
%S_{c} \end{pmatrix}, %\tag{A.4} 
\label{A.4}
\end{equation}
The coupling tensors to other spins $\mathbf{A}^{(1)}(\mathbf{r}_i) \ (i = 2, 3, 4)$ can be derived by
applying symmetry operations that leave the Li site invariant. 
For example, the Li site is on the mirror plane perpendicular to the $b$-axis. Since the Cu1 site is 
transformed to Cu2 by the reflection with respect to this plane, $\mathbf{A}^{(1)}(\mathbf{r}_2)$ is 
given by
\begin{equation}
\mathbf{A}^{(1)}(\mathbf{r}_2) = \begin{pmatrix}
A_{aa}^{(1)} & -A_{ab}^{(1)} & A_{ac}^{(1)} \\
-A_{ba}^{(1)} & A_{bb}^{(1)} & -A_{bc}^{(1)} \\
A_{ca}^{(1)} & -A_{cb}^{(1)} & A_{cc}^{(1)} \end{pmatrix}.
%\tag{A.10} 
\label{A.10}
\end{equation}
Similarly, by considering a twofold rotation with respect to the $b$-axis, which transforms Cu1 to Cu3 
and Cu2 to Cu4, we obtain  
$\mathbf{A}^{(1)}(\mathbf{r}_3) = \mathbf{A}^{(1)}(\mathbf{r}_2)$ and 
$\mathbf{A}^{(1)}(\mathbf{r}_4) = \mathbf{A}^{(1)}(\mathbf{r}_2)$.
Therefore, the contribution from the four nearest neighbor spins to the Fourier components of the coupling 
tensor is expressed as
\begin{equation}
	\begin{split}
\mathbf{A}(\mathbf{q}) &\sim \sum_{i=1}^4 \mathbf{A}^{(1)}(\mathbf{r}_i) e^{i\mathbf{q} \cdot \mathbf{r}_i} \\
&= \begin{pmatrix}
A_{aa}^{(1)} \Theta_1^{(1)}(\mathbf{q}) & A_{ab}^{(1)} \Theta_2^{(1)}(\mathbf{q}) & A_{ac}^{(1)} \Theta_1^{(1)}(\mathbf{q}) \\
A_{ba}^{(1)} \Theta_2^{(1)}(\mathbf{q}) & A_{bb}^{(1)} \Theta_1^{(1)}(\mathbf{q}) & A_{bc}^{(1)} \Theta_2^{(1)}(\mathbf{q}) \\
A_{ca}^{(1)} \Theta_1^{(1)}(\mathbf{q}) & A_{cb}^{(1)} \Theta_2^{(1)}(\mathbf{q}) & A_{cc}^{(1)} \Theta_1^{(1)}(\mathbf{q}) %\tag{A.11} 
\label{A.11}
\end{pmatrix},
	\end{split}
\end{equation}
where the phase factors $\Theta_{1}^{(1)}(\mathbf{q})$, $\Theta_{2}^{(1)}(\mathbf{q})$ are defined as 
\begin{equation}
\begin{split}
\Theta_{1}^{(1)}(\mathbf{q}) &= 4 \cos \left(\frac{q_aa-q_cc}{4}\right) \cos \left(\frac{q_b b}{4}\right) \\
\Theta_{2}^{(1)}(\mathbf{q}) &= 4 \sin \left(\frac{q_aa-q_cc}{4}\right) \sin \left(\frac{q_b b}{4}\right).
\label{LargeTheta}
%A_{\mu\nu}^2(\Theta_1(\mathbf{q}) &= \bigg\{ - 4 A_{\mu\nu}^{(1)} \cos \left(\frac{q_aa-q_cc}{4}\right) \cos \left(\frac{q_b b}{4}\right) \\
%&- 4 A_{\mu\nu}^{(2)} \cos \left(\frac{3 q_aa+q_cc}{4}\right) \cos \left(\frac{q_b b}{4}\right) -\cdots \bigg\}^2 \\
%A_{\mu\nu}^2(\Theta_2(\mathbf{q}) &= \bigg\{ -4 A_{\mu\nu}^{(1)} \sin \left(\frac{q_aa-q_cc}{4}\right) \sin \left(\frac{q_b b}{4}\right) \\
%&- 4 A_{\mu\nu}^{(2)} \sin \left(\frac{3 q_aa+q_cc}{4}\right) \sin \left(\frac{q_b b}{4}\right) -\cdots \bigg\}^2
\end{split}
\end{equation}

In the paramagnetic state the moments are uniform and its Fourier component is given as 
$\langle \mathbf{S}(\mathbf{q}) \rangle = \sqrt{N} \langle \mathbf{S}_0 \rangle \delta(\mathbf{q})$.
Then from Eqs.~\eqref{A.3}, \eqref{A.11}, and \eqref{LargeTheta}, we obtain  
\begin{equation}
\langle \mathbf{H}_\mathrm{hf} \rangle = 4 \begin{pmatrix}
A_{aa}^{(1)} & 0 & A_{ac}^{(1)} \\
0 & A_{bb}^{(1)} & 0 \\
A_{ca}^{(1)} & 0 & A_{cc}^{(1)}
\end{pmatrix} \cdot \mathbf{g} \cdot \langle \mathbf{S}_0 \rangle. %\tag{A.8}
\label{A.8}
\end{equation}
The contribution from the second and further neighbor spins can be added in a straight forward way since 
one $^{7}$Li nucleus always has four $n$-th neighbor spins at an equal distance, which are related by 
the same symmetry operations. The contribution from distant spins can be included by 
replacing  $A_{ii}^{(1)} \ (i = a, b, c)$ in Eq.~\eqref{A.8} with $\sum_n A_{ii}^{(n)}$, 
where $\mathbf{A}^{(n)}$ is the hyperfine coupling tensor to the $n$-th neighbor sites.  
The $K - \chi$ linear relation is obtained by rewriting $\langle S_0 \rangle$ in 
Eq.~\eqref{A.8} by the susceptibility as $\chi = N g \mu_B \langle S_0 \rangle/H_\mathrm{ext}$. 
From Eq.~\eqref{Kchi2}, we obtain 
\begin{equation}
C^\mathrm{tr}_i + C^\mathrm{dip}_i = 4 \sum_n A_{ii}^{(n)} \ (i = a, b, c). \label{CA}
\end{equation}

Similar discussion can be applied to the helically ordered phase with the magnetic wave vector 
$\mathbf{Q}_0$. The Fourier components of spins in the helical phase (and also in the SDW phase) 
are described as $\langle \mathbf{S}(\mathbf{q}) \rangle = \sqrt{N} (\langle \mathbf{S}_Q \rangle \delta(\mathbf{q} - \mathbf{Q}_0) + \langle \mathbf{S}_Q \rangle^* \delta(\mathbf{q} + \mathbf{Q}_0))$.
One $^7$Li nucleus has four nearest neighbor Cu spins located on two chains separated by
$(\mathbf{a}+\mathbf{c})/2$, which are antiferromagnetically coupled, 
$\langle \mathbf{S}_1 \rangle = -\langle \mathbf{S}_3 \rangle$, 
$\langle \mathbf{S}_2 \rangle = -\langle \mathbf{S}_4 \rangle$, since $Q_{0a} = 2\pi$. Therefore,  
one of the phase factor is canceled out, $\Theta_1^{(1)}(\mathbf{Q}_0) = 0$, and the hyperfine field 
is expressed as 

\begin{equation}
\langle \mathbf{H}_\mathrm{hf} \rangle = \begin{pmatrix}
0 & A_{ab}^{(1)} \Theta_2^{(1)}(\mathbf{Q}_0)  & 0 \\
A_{ba}^{(1)} \Theta_2^{(1)}(\mathbf{Q}_0)  & 0 & A_{bc}^{(1)} \Theta_2^{(1)}(\mathbf{Q}_0) \\
0 & A_{cb}^{(1)} \Theta_2^{(1)}(\mathbf{Q}_0)  & 0
\end{pmatrix} \cdot \mathbf{g} \cdot \langle \mathbf{S}_Q \rangle. %\tag{A.12} 
\label{A.12}
\end{equation}

This selection rule obtained from symmetry properties does not change by taking into account 
contribution from distant spins. Here we can consider contribution from four $n$-th neighbor 
Cu spins (n $\geq$ 2) and define the phase factors $\Theta_i^{(n)} (i = 1, 2)$ in a similar manner.
For example, for the second neighbor spins we have 
\begin{equation}
\begin{split}
\Theta_1^{(2)}(\mathbf{q}) &= 4 \cos \left(\frac{3 q_aa+q_cc}{4}\right) \cos \left(\frac{q_b b}{4}\right) \\
\Theta_2^{(2)}(\mathbf{q}) &= 4 \sin \left(\frac{3 q_aa+q_cc}{4}\right) \sin \left(\frac{q_b b}{4}\right).
\label{LargeTheta2}
\end{split}
\end{equation}
Since antiferromagnetic coupling between two neighboring CuO$_2$ chains leads to the cancellation
$\Theta_1^{(2)}(\mathbf{Q}_0) = 0$, the hyperfine field from the second neighbors has the same selection 
rule as Eq.~\eqref{A.12}.

The result in Eq.~\eqref{A.12} indicates that the $a$-component of the ordered moment produces a hyperfine 
field along the $b$-direction. Likewise, the $b$- and $c$-components of the ordered moment produce hyperfine
fields in the $ac$-plane and along the $b$-direction, respectively.
Since the external field is much larger than the hyperfine field ($H_\mathrm{ext} \gg H_\mathrm{hf}$),
only the component of $\mathbf{H}_\mathrm{hf}$ parallel to $\mathbf{H}_\mathrm{ext}$ affects the NMR
line shape. The double-horn line shape of the $^{7}$Li NMR spectra for $H \parallel a$ at low temperatures 
shown in Fig.~\ref{spectra4T}(a), therefore, indicates that the $b$-component of the ordered moments
must have incommensurate modulation. Similarly, the double-horn spectra for  $H \parallel b$ and 
$H \parallel c$ must be caused by incommensurate modulation of the spins in the $ac$-plane and along the 
$b$-direction, respectively. All these observations are consistent with a helical order within the plane
perpendicular to the field at 4~T as concluded from previous experiments\cite{NMR2, neutron1, neutron2}.  

In order to estimate the magnitude of the helical order, we performed simulation of the $^{7}$Li NMR spectra
as follows. First a histogram of hyperfine field was constructed by calculating the dipolar field from 
helically ordered spins lying in the plane perpendicular to the field within a sphere with the radius of 
20~\AA. The histogram is then convoluted with a Gaussian representing broadening due to various imperfections.  
The simulated curves are shown by the red dashed lines and the Gaussian is indicated by the red solid lines 
in Fig.~\ref{spectra4T}(a-c). The simulated curves reproduce the observed spectra at the lowest temperatures
quite well. The magnitude of the helical moment is estimated as 0.25 $\mu_B$ ($H \parallel a$) and 0.34 $\mu_B$ 
($H \parallel b$ and $H \parallel c$), indicating small spin anisotropy.

\subsubsection{$^{51}$V nuclei}
Now we discuss the $^{51}$V NMR spectra at 4~T.
The configuration of nearest neighbor Cu spins around a V nucleus is illustrated in Fig. \ref{crystal}(b).
We use the same definition Eq.~\eqref{A.4} for the hyperfine coupling tensor $\mathbf{A}^{(1)}(\mathbf{r}_1)$
between the $^{51}$V nucleus and the spin at Cu1 site. By considering the mirror operation with respect to the
$ac$-plane, which transforms Cu1 to Cu2, we obtain 
\begin{equation}
\mathbf{A}^{(1)}(\mathbf{r}_2) = \begin{pmatrix}
A_{aa}^{(1)} & -A_{ab}^{(1)} & A_{ac}^{(1)} \\
-A_{ba}^{(1)} & A_{bb}^{(1)} & -A_{bc}^{(1)} \\
A_{ca}^{(1)} & -A_{cb}^{(1)} & A_{cc}^{(1)} \end{pmatrix}. %\tag{A.5} 
\label{A.5} 
\end{equation}
The mirror operation with respect to the $bc$-plane or the twofold rotation around the $c$-axis
lead to the following form of the hyperfine coupling to Cu3 and Cu4,
\begin{equation}
	\begin{split}
\mathbf{A}^{(1)}(\mathbf{r}_3) &= \begin{pmatrix}
A_{aa}^{(1)} & -A_{ab}^{(1)} & -A_{ac}^{(1)} \\
-A_{ba}^{(1)} & A_{bb}^{(1)} & A_{bc}^{(1)} \\
-A_{ca}^{(1)} & A_{cb}^{(1)} & A_{cc}^{(1)} \end{pmatrix}, \\ 
\mathbf{A}^{(1)}(\mathbf{r}_4) &= \begin{pmatrix}
A_{aa}^{(1)} & A_{ab}^{(1)} & -A_{ac}^{(1)} \\
A_{ba}^{(1)} & A_{bb}^{(1)} & -A_{bc}^{(1)} \\
-A_{ca}^{(1)} & -A_{cb}^{(1)} & A_{cc}^{(1)} \end{pmatrix},  
	\end{split}
\label{A.6}
\end{equation}
respectively. The Fourier transform of the coupling tensor is then written as
\begin{equation}
	\begin{split}
\mathbf{A}(\mathbf{q}) &\sim \sum_{i=1}^4 \mathbf{A}^{(1)}(\mathbf{r}_i) e^{i\mathbf{q} \cdot \mathbf{r}_i} \\
&= \begin{pmatrix}
A_{aa}^{(1)} \theta_1^{(1)}(\mathbf{q}) & A_{ab}^{(1)} \theta_2^{(1)}(\mathbf{q}) & A_{ac}^{(1)} \theta_3^{(1)}(\mathbf{q}) \\
A_{ba}^{(1)} \theta_2^{(1)}(\mathbf{q}) & A_{bb}^{(1)} \theta_1^{(1)}(\mathbf{q}) & A_{bc}^{(1)} \theta_4^{(1)}(\mathbf{q}) \\
A_{ca}^{(1)} \theta_3^{(1)}(\mathbf{q}) & A_{cb}^{(1)} \theta_4^{(1)}(\mathbf{q}) & A_{cc}^{(1)} \theta_1^{(1)}(\mathbf{q})
\end{pmatrix}.
	\end{split}
\label{A.7}
\end{equation}
Here we define the phase factors $\theta_i^{(1)}(\mathbf{q}) \ (i = 1, 2, 3, 4)$ as
\begin{equation}
\begin{split}
\theta_1^{(1)}({\mathbf{q}}) = 4 \cos \left(\frac{q_aa}{2}\right) \cos \left(\frac{q_bb}{4}\right) \\
\theta_2^{(1)}({\mathbf{q}}) = 4 \sin \left(\frac{q_aa}{2}\right) \sin \left(\frac{q_bb}{4}\right) \\
\theta_3^{(1)}({\mathbf{q}}) = 4 i \sin \left(\frac{q_aa}{2}\right) \cos \left(\frac{q_bb}{4}\right) \\
\theta_4^{(1)}({\mathbf{q}}) = 4 i \cos \left(\frac{q_aa}{2}\right) \sin \left(\frac{q_bb}{4}\right). \label{theta}
\end{split}
\end{equation}
Similar to the case of $^{7}$Li, it is straightforward to include the dipolar fields from second and further neighbor spins. In the paramagnetic state, only the diagonal components of Eq.~\eqref{A.7} are non-zero. 
The diagonal components $\sum_n A_{ii}^{(n)}$ and the slope of the $K-\chi$ plots $C_i/N \mu_B$ are related
by Eq.~\eqref{CA}.

In the helically ordered state, the magnetic wave vector in the ordered phase is again defined as 
$\mathbf{Q}_0$. One $^{51}$V nucleus has four nearest neighbor Cu spins located on two chains separated by
$\mathbf{a}$, which are ferromagnetically coupled, ($\langle \mathbf{S}_1 \rangle = \langle 
\mathbf{S}_3 \rangle$, $\langle \mathbf{S}_2 \rangle = \langle \mathbf{S}_4 \rangle$). 
This leads to the cancellation of the phase factors,
$\theta_2^{(1)}(\mathbf{Q}_0) = \theta_3^{(1)}(\mathbf{Q}_0) = 0$.
The hyperfine field at $^{51}$V nuclei is obtained by substituting Eq.~\eqref{A.7} and 
$\langle \mathbf{S}(\mathbf{q}) \rangle = \sqrt{N} (\langle \mathbf{S}_Q \rangle \delta(\mathbf{q} -
\mathbf{Q}_0) + \langle \mathbf{S}_Q \rangle^* \delta(\mathbf{q} + \mathbf{Q}_0 ))$ into Eq.~\eqref{A.3} as
\begin{equation}
\langle \mathbf{H}_\mathrm{hf} \rangle = \begin{pmatrix}
A_{aa}^{(1)} \theta_1^{(1)}(\mathbf{Q}_0) & 0 & 0 \\
0 & A_{bb}^{(1)} \theta_1^{(1)}(\mathbf{Q}_0) & A_{bc}^{(1)} \theta_4^{(1)}(\mathbf{Q}_0) \\
0 & A_{cb}^{(1)} \theta_4^{(1)}(\mathbf{Q}_0) & A_{cc}^{(1)} \theta_1^{(1)}(\mathbf{Q}_0)
\end{pmatrix} \cdot \mathbf{g} \cdot \langle \mathbf{S}_Q \rangle. %\tag{A.9} 
\label{A.9}
\end{equation}
This selection rule derived from symmetry properties does not change by including 
contribution from further neighbor spins, similar to the case of $^7$Li nuclei.

Equation~\eqref{A.9} indicates that for $H \parallel a$, only the $a$-component of the ordered moment
affects the $^{51}$V NMR spectrum. Therefore, a helical order perpendicular to the field should not 
cause broadening of the NMR line. The experimental $^{51}$V NMR spectrum for $H \parallel a$ 
shown in Fig.~\ref{spectra4T}(d) exhibits a Gaussian-like shape at the lowest temperature, 
which is much broader than the spectrum at 6~K. This broadening cannot be due to incommensurate modulation of 
the $a$-component, since such modulation should result in a well defined double horn structure. 
The absence of modulation of the $a$-component revealed by $^{51}$V NMR and its presence for the 
$b$-component indicated by $^{7}$Li NMR uniquely point to a helical order at 4~T. 
The Gaussian broadening in the experimental spectrum suggests certain randomness in the interchain 
order, which results in incomplete cancellation of the phase factors. 
This is likely to be caused
by some disorder such as deficiencies of $^7$Li ions. Furthermore, Fig.~\ref{spectra4T}(d) shows  
interesting behavior near $T_N$ that the spectrum has a flat topped line shape resembling a double horn
structure and is broader than the spectrum at the lowest temperature. This indicates that the interchain
correlation near $T_N$ is weaker compared with the correlation well below $T_N$.  

For $H \parallel b$ and $H \parallel c$, broadening of the NMR spectrum can be caused not only by 
modulation of the longitudinal (parallel to the field) component but also by modulation of the transverse (perpendicular to the field) component due to finite off-diagonal element in Eq.~\eqref{A.9}. 
Since anisotropy of magnetic interactions is expected to be small for a spin 1/2 system, it is reasonable 
to assume a helical order for $H \parallel b$ and for $H \parallel c$. 
The double horn spectrum for $H \parallel b$ (Fig.~\ref{spectra4T}e) should then be ascribed to the 
modulation of the $c$-component of the moment. We have performed simulation of the $^{51}$V NMR spectrum
assuming the helical structure with the same magnitude of moment estimated from the analysis of the 
$^{7}$Li spectra, 0.25~$\mu_B$ for $H \parallel a$ and 0.34~$\mu_B$ for $H \parallel b, c$. In the simulation,
both the transferred hyperfine field from the nearest neighbor spins and the dipolar field from all spins 
within a sphere with the radius of 20~\AA \ are taken into account, using the off-diagonal component $A_{bc}^\mathrm{(1)}$ of the transferred hyperfine coupling tensor as a adjustable parameter. 
The best results are shown by the red dashed curves in Fig.~\ref{spectra4T}(e,f). The Gaussian line shape 
used for the convolution are also shown by the red solid curves. The off-diagonal component 
is estimated to be $A_{bc}^\mathrm{(1)}$ = 0.12 $\pm$ 0.02 T/$\mu_B$. In this analysis, 
the $^{51}$V spectra for $H \parallel b$ and $H \parallel c$ should have nearly identical line shape.
However, the experimental spectra are different: a clear double horn structure is observed only for 
$H \parallel b$. The difference cannot be explained by incomplete interchain ordering and the 
reason is still not understood. 

Concluding this subsection, analysis of the $^{7}$Li NMR spectra at 4~T has lead us to conclude an 
incommensurate order with transverse spin modulation for all field direction. The $^{51}$V NMR spectra
for $H \parallel a$ has provided further constraint that the longitudinal component does not show 
incommensurate modulation, pointing uniquely to a helical order. Our conclusion is consistent with 
the results of the previous neutron diffraction and NMR experiments\cite{NMR2, neutron1, neutron2}.

\subsection{SDW state}
\begin{figure*}[t]
\centering
\includegraphics[width=17cm,clip]{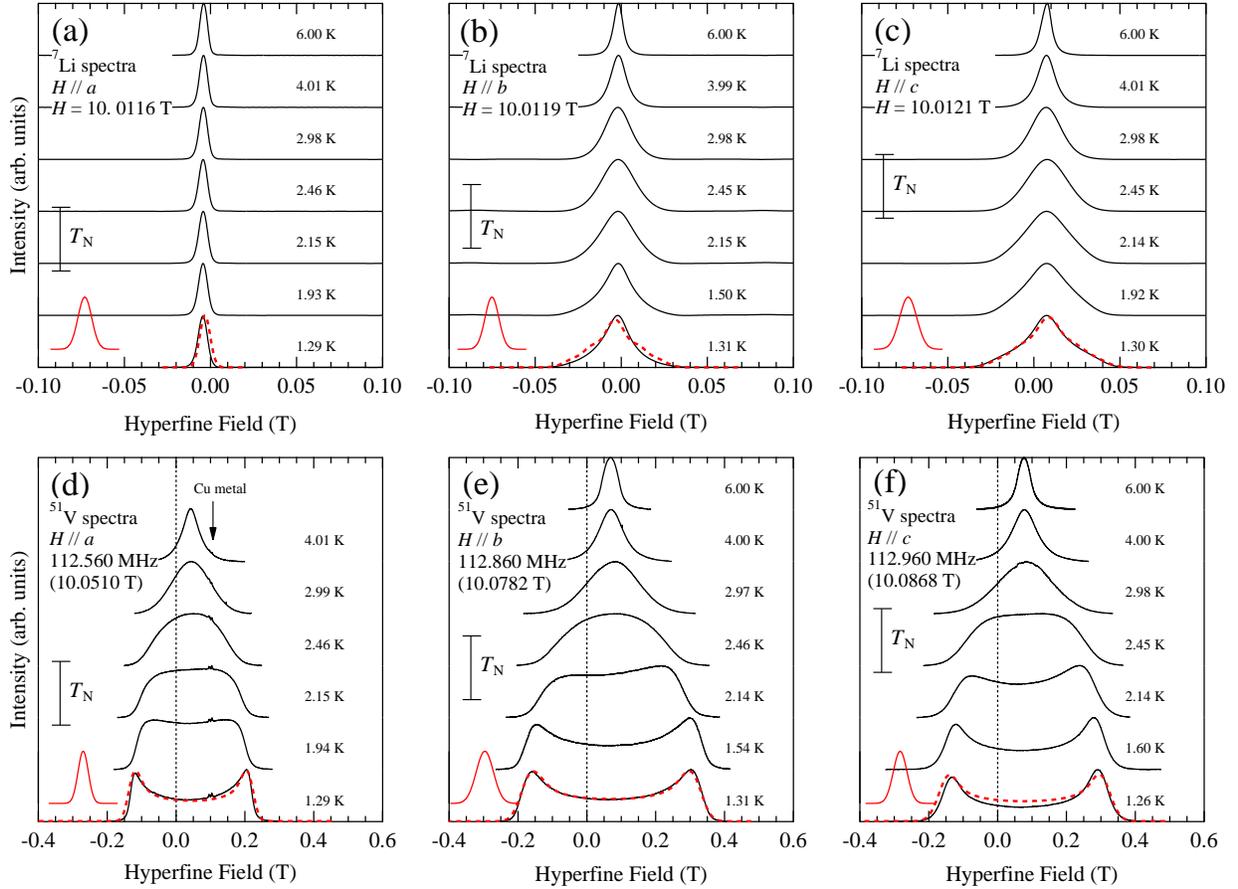}
\caption{ (Color Online) 
Frequency-swept NMR spectra of $^7$Li nuclei for (a) $H \parallel a$, (b) $H \parallel b$, (c) $H \parallel c$
and field-swept NMR spectra of $^{51}$V nuclei for (d) $H \parallel a$, (e) $H \parallel b$ and (f) $H \parallel c$ in the magnetic field near 10 T. The precise values of the fixed magnetic field are indicated in 
Fig. (a-c).  The field values in the parenthesis in Fig. (d-f) correspond to the origin of the 
hyperfine field, $H_0 = \omega/\gamma_N$. Vertical bars denote uncertainties of $T_N$. The red dashed 
curves are simulated spectra for the SDW phase with the moments aligned parallel to a magnetic field.
Short range interchain correlation modeled by a Gaussian distribution of the relative phase of SDW between 
two layers along the $c$-direction is considered in the simulation. The red solid lines in the left side 
of the figures show the Gaussian used for the convolution to simulate the spectra.}
\label{spectra10T}
\end{figure*}
NMR spectra in the magnetic field of 10 T are shown in Fig.~\ref{spectra10T}. The magnetic transition 
temperature $T_N$ is determined by inspecting the change of the $^{51}$V NMR line shape from a double-horn 
structure below  $T_N$ to a single-peak structure above $T_N$. The values of $T_N$ at 10~T thus determined 
are slightly lower than those reported by previous NMR study\cite{NMR4}. A collinear SDW order has been 
concluded by the previous NMR\cite{NMR2, NMR3, NMR4} and neutron diffraction 
experiments\cite{neutron2, neutron3}. The magnetic structure for $H \parallel c$ is illustrated by the 
arrows in Fig.~\ref{crystal}(a). Longitudinal spin component has incommensurate modulation along the chain, 
while the transverse components do not exhibit any ordering. Spins on neighboring CuO$_2$ chains are 
coupled antiferromagnetically in the SDW phase as well as in the helical phase. 
The previous studies suggested that interchain correlation is weak and only short range correlation 
develops along the $c$-direction\cite{NMR4, neutron3}.

\subsubsection{$^7$Li nuclei}
We apply the same analysis as have been done in the helical phase to the NMR spectra in the SDW phase.
The magnetic wave vector in the SDW phase is described as $\mathbf{Q}_0 = 2 \pi(1, q_b, 0)$,
where $q_b$ decreases as the magnetization increases\cite{neutron2, neutron3}.
We can use the same form of the hyperfine coupling tensor in Eq.~\eqref{A.12} at 10 T 
since spin chains order antiferromagnetically ($Q_{0a} = 2 \pi$) in both the helical and SDW phases.
For example, only the $b$-component of spins influences the $^{7}$Li line shape for $H \parallel a$.  
The $^7$Li NMR spectra in Fig.~\ref{spectra10T}(a) show a single-peaked structure with no shift nor 
broadening down to the lowest temperature. Therefore, we conclude that the $b$-component of the 
ordered moment is zero. The NMR spectra for $H \parallel b$ and $H \parallel c$ shown in 
Fig.~\ref{spectra10T}(b) and (c) also have a single peaked structure at low temperatures, 
indicating the absence of the transverse spin components consistent with the SDW order. 
However, the spectra for $H \parallel b$ and $H \parallel c$ are much broader than the spectrum for 
$H \parallel a$ below $T_N$.

The broad line shape of the spectra for $H \parallel b$ and $H \parallel c$ can be explained by 
short range interchain correlation along the $c$-direction. The red dashed curves in Fig.~\ref{spectra10T} 
show simulated spectra for the SDW state with incomplete interchain ordering, which is obtained as 
follows. We first assume complete ferromagnetic correlation for two chains neighboring along the 
$a$-directions. The magnitude of the SDW modulation is determined from the $^{51}$V NMR spectra 
as described below. The short range correlation is modeled by a Gaussian 
distribution in the relative phase of the SDW order for the two neighboring layers displaced by 
$\mathbf{c}/2$. The experimental line shape is well explained by this model when the standard deviation 
of this phase distribution is $2 \pi \times 0.16$. This phase randomness corresponds to the 
spin correlation length equal to the unit cell length along the $c$-direction.
Our analysis is consistent with the observation reported in the earlier NMR and neutron 
experiments\cite{NMR4, neutron3}.

\subsubsection{$^{51}$V nuclei}
The $^{51}$V NMR spectra at 10~T shown in Fig.~\ref{spectra10T}(d-f) exhibit well defined double horn 
structure. We can apply the same analysis as was done for the spectra at 4 T.
The hyperfine field is expressed again by Eq.~\eqref{A.9}. 
Since only the $a$-component of the ordered moment affects the spectra for $H \parallel a$,  
the double-horn line shape for $H \parallel a$ indicates the modulation of the longitudinal component.
For $H \parallel b$ and $H \parallel c$, both the transverse and longitudinal spin components 
contribute to the broadening. However, the transverse modulation has been already ruled out from 
the analysis of the $^7$Li NMR spectra. Therefore, our results uniquely point to the incommensurate 
SDW order with longitudinal modulation for all field directions. 

Simulated NMR spectra in the SDW phase are shown by the red dashed curves in Fig.~\ref{spectra10T}(d-f). 
These are obtained by convolving the calculated histogram of the hyperfine field with a Gaussian broadening functions, which are shown by the red solid curves. We have succeeded to reproduce the experimental spectra
at the lowest temperature quite well by assuming the following spin structure, 
$M_a = (0.10 + 0.45 \sin q_b) \ \mu_B$,
$M_b = (0.11 + 0.48 \sin q_b) \ \mu_B$ and
$M_c = (0.14 + 0.62 \sin q_b) \ \mu_B$, for $H \parallel a$, $H \parallel b$ and $H \parallel c$, respectively.
The anisotropy in the magnitude of the SDW modulation agrees with the anisotropy of the $g$ - factor ($g_{cc} > g_{bb} > g_{aa}$)\cite{ESR}. The width of the Gaussians is nearly same as the width of the spectra at 6 K. This width, however, is broader that the overall quadrupole splitting, indicating that it is dominated by magnetic inhomogeneity. Also the spectra in the paramagnetic phase near $T_N$ is much broader than the Gaussian used for convolution, indicating enhanced effects of disorder near $T_N$. We have also examined how the line shape is 
influenced by the incomplete interchain ordering. Compared with the case of $^7$Li spectra, however, interchain randomness does not influence the $^{51}$V spectra strongly.

\subsection{Transition between the helical and SDW states}
\begin{figure}[b]
\centering
\includegraphics[width=7cm,clip]{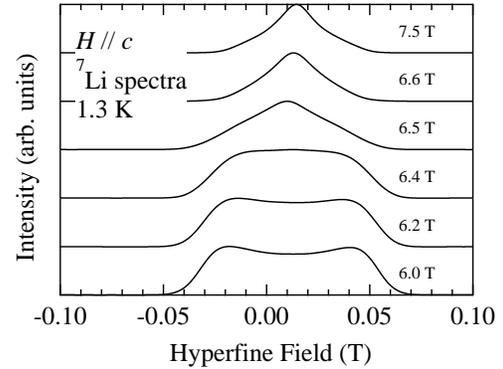} 
\caption{
Field dependencies of $^7$Li NMR spectra for $H \parallel c$ in 1.3 K. 
No hysteresis was observed with respect to the magnetic field.}
\label{spectraLiT}
\end{figure}
Figure \ref{spectraLiT} shows the variation of $^7$Li NMR spectra across the transition between the helical and 
SDW states. By comparing these NMR spectra with the ones at 4~T and 10~T, we can conclude that the ground state 
has a helical order below 6.2 T and the SDW order above 6.6~T. The line shape changes rapidly between 6.2~T 
and 6.6~T. The sharp magnetic transition is consistent with that observed in the magnetization curve at 
1.3~K\cite{magnetization}. In a narrow field range of 6.4 - 6.5~T, the NMR spectra indicate coexistence of 
two phases, while no clear hysteresis with respect to the magnetic field was observed. 
These results indicate that the magnetic transition between the helical and the SDW phases is weakly 
first-order, as is generally expected from distinct symmetries of the two phases.
In the previous neutron diffraction experiments, coexistence of the Bragg peaks from two phases 
was observed at much higher field of 7.8~T\cite{neutron3} and in a wider field range 
8.5 - 9~T\cite{neutron2}. The sample dependence in the transition behavior may be related to the amount of 
disorder such as Li-deficiency. 

\section{Spin lattice relaxation rate $1/T_1$}
\begin{figure*}[t]
\centering
\includegraphics[width=17cm,clip]{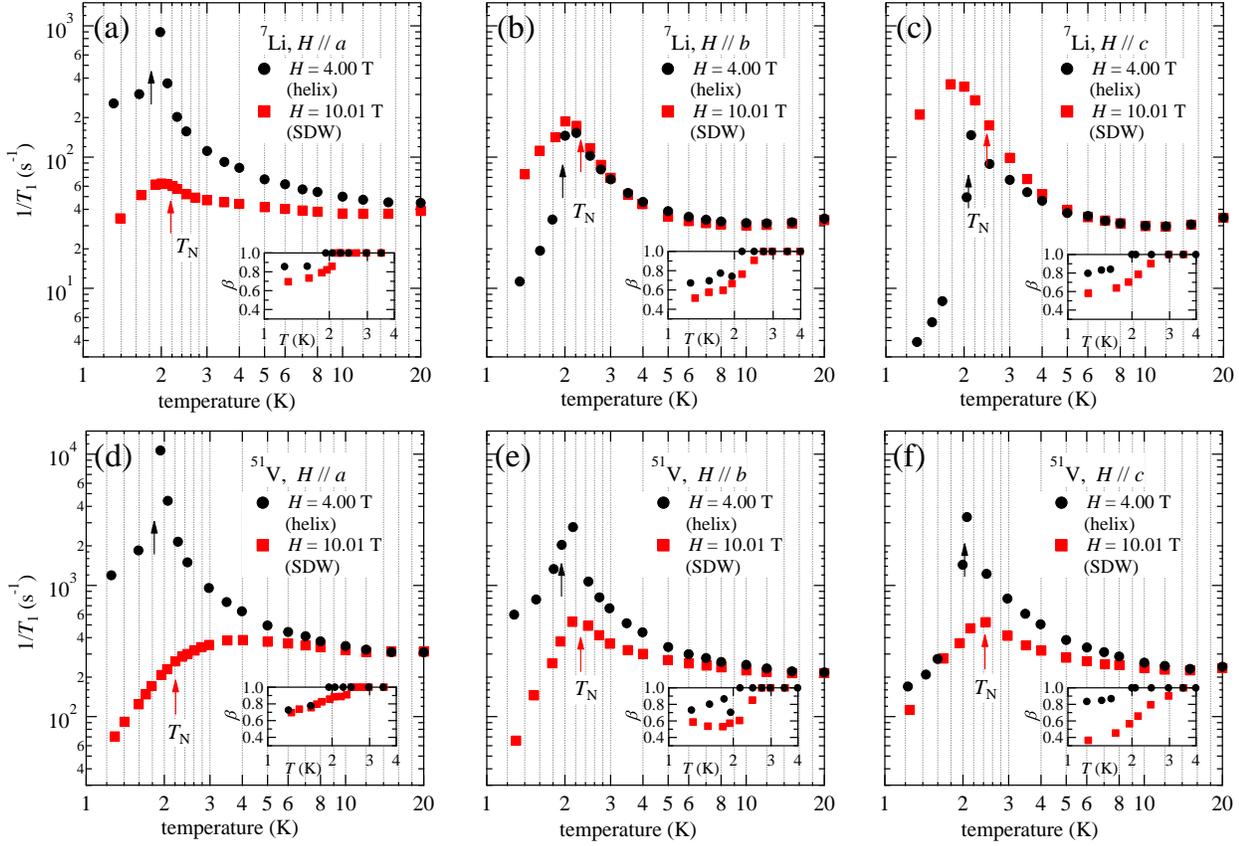}
\caption{(Color Online) 
Temperature dependences of $1/T_1$ for $^7$Li nuclei (a-c) and $^{51}$V nuclei (d-f). 
The solid circles show $1/T_1$ measured in a magnetic field of 4~T, where a helical spin order appears 
below $T_N$, and the solid squares show the data at 10~T, where SDW phase appears below $T_N$.
The arrows indicate the transition temperature $T_N$ determined from the broadening of the 
NMR spectrum. The insets show the temperature dependence of the stretch exponent $\beta$,
which provides a measure of the inhomogeneous distribution of $1/T_1$.}
\label{T1}
\end{figure*}
In this section, we present the results of nuclear spin-lattice relaxation rate $1/T_1$ at both  
$^7$Li and $^{51}$V nuclei and quantitatively discuss the anisotropic spin fluctuations. 
A general formula of $1/T_1$ is given as 
\begin{equation}
\frac{1}{T_1} = \frac{\gamma_N^2}{2} \int dt e^{i\omega t} \langle \{ H_\mathrm{hf}^+(t), H_\mathrm{hf}^-(0) \} \rangle, \label{T1eq}
\end{equation}
where $\omega$ is the NMR frequency, $\langle \rangle$ indicates the thermal average, and 
$\{ A, B \} \equiv (AB + BA)/2$. The transverse components of hyperfine field is defined as  
\begin{equation}
H_\mathrm{hf}^\pm \equiv H_\mathrm{hf, \mu} \pm i H_\mathrm{hf, \nu},
\end{equation}
where $\mu$ and $\nu$ denote the two orthogonal directions perpendicular to the magnetic field. 
By substituting Eq.~\eqref{A.3} representing the hyperfine field in terms of the Fourier components
of spins into Eq,~\eqref{T1eq}, $1/T_1$ for the field along the $\xi$-direction, $(1/T_1)_\xi$ ($\xi = a, b, c$),
can be written as 

\begin{equation}
\begin{split}
\left( \frac{1}{T_1} \right)_a &= \frac{\gamma_N^2}{2N} \sum_\mathbf{q} 
\bigg[ \left\{ |A(\mathbf{q})_{bb}|^2 + |A(\mathbf{q})_{bc}|^2 \right\} g_{bb}^2 S_{bb}(\mathbf{q}, \omega) \\ 
&+ \left\{ |A(\mathbf{q})_{bc}|^2 + |A(\mathbf{q})_{cc}|^2 \right\} g_{cc}^2 S_{cc}(\mathbf{q}, \omega) \\
&+ \left\{ |A(\mathbf{q})_{ab}|^2 + |A(\mathbf{q})_{ac}|^2 \right\} g_{aa}^2 S_{aa}(\mathbf{q}, \omega) \bigg] \\
\left( \frac{1}{T_1} \right)_b &= \frac{\gamma_N^2}{2N} \sum_\mathbf{q} 
\bigg[ \left\{ |A(\mathbf{q})_{cc}|^2 + |A(\mathbf{q})_{ac}|^2 \right\} g_{cc}^2 S_{cc}(\mathbf{q}, \omega) \\ 
&+ \left\{ |A(\mathbf{q})_{ac}|^2 + |A(\mathbf{q})_{aa}|^2 \right\} g_{aa}^2 S_{aa}(\mathbf{q}, \omega) \\
&+ \left\{ |A(\mathbf{q})_{ab}|^2 + |A(\mathbf{q})_{bc}|^2 \right\} g_{bb}^2 S_{bb}(\mathbf{q}, \omega) \bigg] \\
\left( \frac{1}{T_1} \right)_c &= \frac{\gamma_N^2}{2N} \sum_\mathbf{q} 
\bigg[ \left\{ |A(\mathbf{q})_{aa}|^2 + |A(\mathbf{q})_{ab}|^2 \right\} g_{aa}^2 S_{aa}(\mathbf{q}, \omega) \\ 
&+ \left\{ |A(\mathbf{q})_{ab}|^2 + |A(\mathbf{q})_{bb}|^2 \right\} g_{bb}^2 S_{bb}(\mathbf{q}, \omega) \\
&+ \left\{ |A(\mathbf{q})_{ac}|^2 + |A(\mathbf{q})_{bc}|^2 \right\} g_{cc}^2 S_{cc}(\mathbf{q}, \omega) \bigg],
\label{commonT1}
\end{split}
\end{equation}  
where the dynamical spin correlation function $S_{\mu \nu}(\mathbf{q}, \omega)$ is defined as 
\begin{equation}
S_{\mu \nu}(\mathbf{q}, \omega) 
\equiv \int^\infty_{-\infty} dt e^{i\omega t} \langle \{ S(\mathbf{q})_{\mu}(t), S(-\mathbf{q})_{\mu}(0) \} \rangle, \label{corr}
\end{equation}
and the following relation ensured by the crystal symmetry is used,  
\begin{equation}
\langle \{ S(\mathbf{q})_{\mu}(t), S(\mathbf{q}')_{\nu}(0) \} \rangle = \langle \{ S(\mathbf{q})_{\mu}(t), S(-\mathbf{q})_{\mu}(0) \} \rangle 
\delta_{\mathbf{q}\mathbf{-q}'} \delta_{\mu\nu}. \label{spin}
\end{equation}

The relaxation rate $1/T_1$ of $^7$Li nuclei and $^{51}$V nuclei is plotted against temperature
in Fig.~\ref{T1}(a-c) and (d-f), respectively. As a general trend, $1/T_1$
shows a peak near $T_N$ due to slowing down of the spin fluctuations towards the 
spin order. Here the ordering temperature $T_N$ is determined from the variation of the 
$^7$Li NMR spectrum at 4~T and of the $^{51}$V NMR spectrum at 10~T.
At 10 T, the peak temperature of $1/T_1$ for $^7$Li is lower than $T_N$
while the peak of $1/T_1$ coincides with $T_N$ for $^{51}$V. The reason is not understood.
The temperature dependence of the stretch exponent $\beta$ is also shown in the inset of Fig.~\ref{T1}.
Recovery curves are fit well to the single exponential function above $T_N$, while we had to use stretched exponential function below $T_N$. This is likely to be caused by spatial distribution of $1/T_1$ due to the incommensurate spin ordering below $T_N$.

The peak of $1/T_1$ near $T_N$ extends over a wide temperature range. The increase of $1/T_1$ starts as high 
as 20~K, which is an order of magnitude higher than $T_N$. Such a wide temperature range of short range
correlation is often ascribed to 1D fluctuations, which are described in the framework of Tomonaga-Luttinger (TL) liquid. In fact, theories have predicted power law temperature dependence for $1/T_1$ in 1D frustrated chains with the exponent determined by the TL parameters\cite{1DtheoryofT11,1DtheoryofT12}. However, it is 
difficult for the data shown in Fig.~\ref{T1} to find a wide enough temperature range, over which the power 
law behavior can be established. It is likely that the interchain interaction in $\mathrm{LiCuVO_4}$ is not sufficiently weak. Hence we do not attempt to analyze the $1/T_1$ data in terms of TL liquids.        

\subsection{$1/T_1$ of $^7$Li nuclei}
Although $1/T_1$ of $^7$Li nuclei always shows a peak near $T_N$, the magnitude of the peak is strongly 
anisotropic and the anisotropy changes with magnetic field. At 4~T, where the helical order appears at low
temperatures, the peak is most pronounced for $H \parallel a$. However, the peak is weakest for 
$H \parallel a$ at 10~T, where the SDW order is stabilized. In order to understand such field dependent 
anisotropy, we present semi-quantitative analysis of $1/T_1$. 

In the following discussion, we focus on the values of $1/T_1$ at $T_N$, where the spin fluctuations should be 
strongly enhanced only in a very narrow region in $q$-space around the ordering wave vector $\mathbf{Q}_0$. 
Then the $q$-dependent hyperfine coupling constants in Eq.~\eqref{commonT1} can be replaced by their values 
at $\mathbf{Q}_0$ and taken out of the $q$-sum. What remains is the average of the spin correlation function 
over $\mathbf{q}$, which we denote $\langle S(\mathbf{q}, \omega) \rangle$. It is dominated by the sharp peak of 
$S(\mathbf{q}, \omega)$ at $\mathbf{Q}_0$.  

As indicated in Eq.~\eqref{A.12}, certain components of the hyperfine coupling tensor of $^7$Li nuclei 
vanish at $\mathbf{Q}_0$ for symmetry reasons. This selection rule remains valid when the long 
range dipolar interaction is taken into account. Hence,
\begin{equation}
\mathbf{A}(\mathbf{Q}_0)  
= \begin{pmatrix}
0 & A_{ab}(\mathbf{Q}_0) & 0 \\
A_{ba}(\mathbf{Q}_0) & 0 & A_{bc}(\mathbf{Q}_0) \\
0 & A_{cb}(\mathbf{Q}_0) & 0 
\label{Litensor}
\end{pmatrix}.
\end{equation}
The general expression Eq.~\eqref{commonT1} can then be simplified for $^7$Li nuclei as 
\begin{equation}
\begin{split}
\left( \frac{1}{T_1} \right)_a & = \frac{\gamma_N^2}{2} 
\bigl\{  A_{bc}(\mathbf{Q}_0)^2 \ (g_{bb}^2 + g_{cc}^2) \ \langle S_\perp (\mathbf{q}, \omega) \rangle \\
&+  A_{ab}(\mathbf{Q}_0)^2 \ g_{aa}^2 \ \langle S_\parallel (\mathbf{q}, \omega) \rangle \bigr\} \\
\left( \frac{1}{T_1} \right)_b & = \frac{\gamma_N^2}{2}
\bigl( A_{bc}(\mathbf{Q}_0)^2 + A_{ab}(\mathbf{Q}_0)^2 \bigr) \ g_{bb}^2 \ 
\langle S_\parallel (\mathbf{q}, \omega) \rangle \\
\left( \frac{1}{T_1} \right)_c & = \frac{\gamma_N^2}{2}
\big\{ A_{ab}(\mathbf{Q}_0)^2 \ (g_{aa}^2 + g_{bb}^2) \ \langle S_\perp (\mathbf{q}, \omega) \rangle \\
&+  A_{bc}(\mathbf{Q}_0)^2 \ g_{cc}^2 \ \langle S_\parallel (\mathbf{q}, \omega) \rangle  \big\}. \label{LiT1}
\end{split}
\end{equation}
Here $\langle S_\perp (\mathbf{q}, \omega) \rangle$ and $\langle S_\parallel (\mathbf{q}, \omega) \rangle$ stand for the transverse 
and longitudinal spin correlation functions with respect to the field. For the case of $H \parallel a$,
for example, $\langle S_\perp (\mathbf{q}, \omega) \rangle = \langle S_{bb} (\mathbf{q}, \omega) \rangle = \langle S_{cc} (\mathbf{q}, \omega) \rangle$ 
and $\langle S_\parallel (\mathbf{q}, \omega) \rangle = \langle S_{aa}(\mathbf{q}, \omega) \rangle$. Since anisotropic response of a 
spin 1/2 Heisenberg system such as $\mathrm{LiCuVO_4}$ is supposed to be caused solely by external magnetic field,  $\langle S_\perp (\mathbf{q}, \omega) \rangle$ and $\langle S_\parallel (\mathbf{q}, \omega) \rangle$ should not depend on the 
field direction except for minor effects due to anisotropy of the g-value. We assume this in the following, i.e.     
$\langle S_{aa}(\mathbf{q}, \omega) \rangle$ and $\langle S_{bb} (\mathbf{Q}_0, \omega) \rangle = \langle S_{cc} (\mathbf{q}, \omega) \rangle$ for 
$H \parallel a$ are equal to $\langle S_{bb}(\mathbf{q}, \omega) \rangle$ and $\langle S_{cc} (\mathbf{q}, \omega) \rangle = 
\langle S_{aa} (\mathbf{q}, \omega) \rangle$ for $H \parallel b$, respectively, and also equal to 
$\langle S_{cc}(\mathbf{q}, \omega) \rangle$ and $\langle S_{aa} (\mathbf{q}, \omega) \rangle = \langle S_{bb} (\mathbf{q}, \omega) \rangle$ for $H \parallel c$.  Then the relaxation rate for different field directions can be expressed in a unified way,
\begin{equation}
\left( \frac{1}{T_1} \right)_\xi = \Gamma^{\perp}_\xi \langle S_\perp (\mathbf{q}, \omega) \rangle 
+ \Gamma^{\parallel}_\xi \langle S_\parallel (\mathbf{q}, \omega) \rangle \ \ \ (\xi = a, b, c), 
\label{LiT1uni}
\end{equation}
where the definitions of the coefficients $\Gamma^{\perp}_\xi$ and $\Gamma^{\parallel}_\xi$ are obvious
by comparing Eqs.~\eqref{LiT1} and \eqref{LiT1uni}. 

The values of $\Gamma^{\perp}_\xi$ and $\Gamma^{\parallel}_\xi$ for $^7$Li nuclei for the field of 4 and 
10~T are listed in the upper part of Table  \ref{coeff}. They are calculated as follows. First, since the hyperfine field at $^7$Li nuclei is solely due to the dipolar interaction, the values of the tensor components
$A_{\mu \nu}$ in Eq.~\eqref{Litensor} can be calculated by lattice sum within a sphere with the radius of 
60~\AA \ from a $^7$Li nucleus. The ordering wave vector $\mathbf{Q}_0$ in the helical phase does not change 
with field, therefore, we can use the zero field value $\mathbf{Q}_0 = 2 \pi (1, 0.468, 0)$ at 4~T for all 
field directions. In the SDW phase, the incommensurate wave vector along the chain is related to the 
magnetization by $q_b = \pi(1/2 - \langle S_z \rangle)$, as was confirmed by the neutron diffraction experiments\cite{neutron2, neutron3}. Then we can determine $\mathbf{Q}_0$ at 10~T from the magnetization
data as $\mathbf{Q}_0 = 2 \pi (1, 0.442, 0)$ ($H \parallel a$ and $b$) and
$\mathbf{Q}_0 = 2 \pi (1, 0.435, 0)$ ($H \parallel c$). The diagonal components of $g$-tensor were 
determined by the electronic spin resonance measurements\cite{ESR} as $g_{aa} = 2.070$, $g_{bb} = 2.095$ 
and $g_{cc} = 2.313$. 

The experimental data of $1/T_1$ of $^7$Li nuclei at $T_N$ are also listed in Table \ref{coeff}.
The anisotropy of $1/T_1$ of $^7$Li changes drastically with magnetic field. At 4~T it is by far the largest
for $H \parallel a$ and nearly an order of magnitude smaller for other directions. This anisotropy agrees
largely with the anisotropy of $\Gamma^{\perp}$ but not with $\Gamma^{\parallel}$. This indicates that 
$\langle S_\perp (\mathbf{q}, \omega) \rangle \gg \langle S_\parallel (\mathbf{q}, \omega) \rangle$,
i.e. the transverse fluctuations are dominant at 4~T.  The anisotropy becomes opposite at 10~T; $1/T_1$
gets smallest for $H \parallel a$. This agrees with the anisotropy of $\Gamma^{\parallel}$,
indicating that the longitudinal fluctuations become dominant. Since the transverse helical order
and the longitudinal SDW order are stabilized at low temperatures at 4~T and 10~T, respectively, these 
results are indeed quite natural because fluctuations of the order parameter should be dominant near 
$T_N$.   

Let us now evaluate the field dependence of the spin correlation function more quantitatively. 
For this purpose, the results for $H \parallel b$ deserve particular attention since $\Gamma^{\perp}_b$ = 0
and only the longitudinal fluctuations contribute to $1/T_1$ at $^7$Li sites. The temperature dependence of 
$1/T_1$ for $H \parallel b$ (Fig.~\ref{T1}b) shows a clear peak near $T_N$. This indicates that the 
longitudinal spin fluctuations grow with decreasing temperature even at 4~T, where the transverse helical 
order is eventually stabilized. The peak in $1/T_1$ increases only slightly when the field is increased from 
4 to 10~T, indicating, rather unexpectedly, that the longitudinal fluctuations are not enhanced much 
at 10~T even though they do correspond to the order parameter at low temperatures. Since 
$(1/T_1)_b = \Gamma^{\parallel}_b \langle S_\parallel (\mathbf{q}, \omega) \rangle$ for $H \parallel b$
we can immediately evaluate $\langle S_\parallel (\mathbf{q}, \omega) \rangle$ from the experimental data of 
$(1/T_1)_b$, which is plotted in Fig.~\ref{SF} by the open circles. 

We should remark that our analysis is based on somewhat oversimplified 
assumptions. In particular, spin correlation is assumed to have very sharp peak at $\mathbf{Q}_0$. 
Although we expect the peak to be sharp along the chain ($b$-direction), this may not be so along the 
$c$-direction. In fact, the analysis of the $^7$Li €NMR spectrum at 10~T indicates that antiferromagnetic 
correlation between two nearest neighbor chains are only about 60~\%. Therefore, we expect that our analysis 
of $1/T_1$ has at most semi-quantitative validity.

\subsection{$1/T_1$ of $^{51}V$ nuclei}
The temperature dependence of $1/T_1$ of $^{51}$V nuclei shows most spectacular change of behavior 
with magnetic field for $H \parallel a$ as shown in Fig.~\ref{T1}(d). At 4~T, $(1/T_1)_a$ exhibits 
strong enhancement with decreasing temperature over a wide temperature range above $T_N$ followed by 
a pronounced peak near $T_N$. At 10~T, on the other hand, it keeps decreasing with decreasing 
temperature starting far above $T_N$ and no peak nor anomaly can be detected near $T_N$. In fact,
the data at 10~T can be fitted to an exponential temperature dependence with the activation energy of 
1.8~K, suggesting an energy gap in the spin excitation spectrum. For other field directions, $1/T_1$
shows a peak near $T_N$ at 10~T, although it is much weaker than the peak at 4~T. In order to understand
such behavior, we perform similar semi-quantitative analysis as was done for $^7$Li nuclei.

For $^{51}$V nuclei, the transferred hyperfine field from nearest neighbor spins is an order of magnitude 
larger than the dipolar field from further neighbors. Therefore, it is sufficient for quantitative 
analysis of $1/T_1$ to consider the hyperfine coupling to the four nearest neighbor spins. 
Following the same approximation as was assumed for the analysis of $^{7}$Li results, 
the hyperfine coupling tensor at $\mathbf{Q}_0$ is given by  
\begin{equation}
\mathbf{A}(\mathbf{Q}_0)  
= \begin{pmatrix}
A_{aa}(\mathbf{Q}_0) & 0 & 0 \\
0 & A_{bb}(\mathbf{Q}_0) & A_{bc}(\mathbf{Q}_0) \\
0 & A_{cb}(\mathbf{Q}_0) & A_{cc}(\mathbf{Q}_0) 
\label{VtensorforT1} 
\end{pmatrix}, 
\end{equation}
and $1/T_1$ is given as 
\begin{equation}
\begin{split}
\left( \frac{1}{T_1} \right)_a & = \frac{\gamma_N^2}{2} 
\bigl\{ A_{bb}(\mathbf{Q}_0)^2 g_{bb}^2 + A_{cc}(\mathbf{Q}_0)^2 g_{cc}^2 \\
&+ A_{bc}(\mathbf{Q}_0)^2 (g_{bb}^2 + g_{cc}^2) \ \bigr\} \ \langle S_\perp (\mathbf{q}, \omega) \rangle \\
\left( \frac{1}{T_1} \right)_b & = \frac{\gamma_N^2}{2}
\big\{ \bigl( A_{cc}(\mathbf{Q}_0)^2 g_{cc}^2 + A_{aa}(\mathbf{Q}_0)^2 g_{aa}^2 \bigr) \
\langle S_\perp (\mathbf{q}, \omega) \rangle \\
&+ A_{bc}(\mathbf{Q}_0)^2 g_{bb}^2 \ \langle S_\parallel (\mathbf{q}, \omega) \rangle \big\}  \\
\left( \frac{1}{T_1} \right)_c & = \frac{\gamma_N^2}{2}
\big\{ \bigl( A_{aa}(\mathbf{Q}_0)^2 g_{aa}^2 + A_{bb}(\mathbf{Q}_0)^2 g_{bb}^2 \bigr) \
\langle S_\perp (\mathbf{q}, \omega) \rangle \\
&+ A_{bc}(\mathbf{Q}_0)^2 g_{cc}^2 \ \langle S_\parallel (\mathbf{q}, \omega) \rangle \big\},  
\label{VT1}
\end{split} 
\end{equation}
which can be expressed in the same formula of Eq.~\eqref{LiT1uni} as in the case of $^{7}$Li nuclei.

The values of the coefficients $\Gamma^{\perp}_\xi$ and $\Gamma^{\parallel}_\xi$ for $^{51}$V nuclei 
are listed in the lower part of Table \ref{coeff} together with the experimental data of $1/T_1$
at $T_N$. To evaluate $\Gamma^{\perp}_\xi$ and $\Gamma^{\parallel}_\xi$, we used the values of the 
diagonal components of the hyperfine coupling tensor determined from the $K-\chi$ plot as well as the 
off-diagonal component $A_{bc}^{(1)} = 0.12$~T/$\mu_B$ determined from the fitting of the $^{51}$V
NMR spectrum in the helical state. The experimental results of $1/T_1$ show distinct anisotropy for
different magnetic field. At 4~T it is by far the largest for $H \parallel a$ consistent with the 
anisotropy of $\Gamma^{\perp}$. However, at 10~T, it is smallest for $H \parallel a$ in agreement with the anisotropy of $\Gamma^{\parallel}$. These observations go parallel with the results on  $^{7}$Li nuclei
and indicate crossover of the dominant spin fluctuations from transverse to longitudinal modes as the
ground state changes from helical to SDW states with magnetic field.  

\begin{figure}[t]
\centering
\includegraphics[width=8cm,clip]{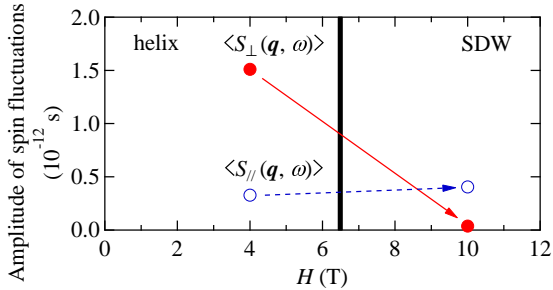}
\caption{(Color Online) 
Field dependences of $\langle S_\perp(\mathbf{q}, \omega) \rangle$ and 
$\langle S_\parallel(\mathbf{q}, \omega) \rangle$
at $T_N$. A straight bold line at 6.5 T describes the first-order phase transition between the helical phase and the SDW phase.}
%$\mathbf{Q}_0$ expresses the magnetic wave vector of the ordered phases.
\label{SF}
\end{figure}

We notice from Eq.~\eqref{VT1} that only the transverse spin fluctuations contribute to $1/T_1$ at 
$^{51}$V site for $H \parallel a$. Therefore, the results in Fig.~\ref{T1}(d) demonstrate remarkably 
contrasting behavior of low frequency transverse spin dynamics. In particular, the activated temperature
dependence of $1/T_1$ at 10~T provides a direct evidence for an energy gap in the transverse spin excitation
spectrum. This is consistent with the theoretical prediction that bound magnon pairs are formed in the 
field range where the SDW correlation becomes dominant over the helical correlation \cite{1Dtheory0, 1Dtheory2, 1Dtheory3}. From the experimental 
data of $1/T_1$ for  $H \parallel a$ we can evaluate $\langle S_\perp (\mathbf{q}, \omega) \rangle$ 
at $T_N$ as plotted in Fig.~\ref{SF} by the solid circles. Again this demonstrates strong suppression 
of the transverse spin fluctuations as the field increases across the boundary between the helical and SDW 
phases.  

\begin{table}
\caption{Coefficients for the contributions from transverse ($\Gamma^\mathrm{\perp}_\xi$) and
longitudinal ($\Gamma^\mathrm{\parallel}_\xi$) spin fluctuations to $1/T_1$.
Also the  experimental values of the spin-relaxation rate ($(1/T_1)_{\mathrm{exp}, \xi}$) are 
compared with the calculated values ($(1/T_1)_{\mathrm{cal}, \xi}$) as described in the text.
At 4~T (10~T), the anisotropy of $(1/T_1)_\mathrm{exp}$ is 
largely in agreement with the anisotropy of $\Gamma^\mathrm{\perp}$ ($\Gamma^\mathrm{\parallel}$) 
as indicated by the numbers in bold face.}

\label{coeff}
\begin{center}
\begin{tabular}{ccccc}
$^7$Li & & & & \\ \hline 
$\xi$ & $\Gamma^\mathrm{\perp}_\xi$ & $\Gamma^\mathrm{\parallel}_\xi$ & $(1/T_1)_{\mathrm{exp}, \xi}$ & $(1/T_1)_{\mathrm{cal}, \xi}$ \\
&  ($10^{12} s^{-2}$) & ($10^{12} s^{-2}$) & (s$^{-1}$) & (s$^{-1}$) \\ \hline
4 T & & & & \\ \hline 
$a$ & $\mathbf{805.3}$ & 79.43 & 897 & 1242 \\
$b$ & $\mathbf{0}$ & 444.3 & 145 & - \\
$c$ & $\mathbf{160.8}$ & 442.4 & 147 & 387.5 \\ \hline
10 T & & & & \\ \hline 
$a$ & 778.8 & $\mathbf{73.88}$ & 60.2 & 58.53 \\
$b$ & 0 & $\mathbf{426.8}$ & 173 & - \\
$c$ & 148.6 & $\mathbf{422.1}$ & 175 & 176.8 \\
\hline \\
$^{51}$V & & & & \\ \hline
$\xi$ & $\Gamma^\mathrm{\perp}_\xi$ & $\Gamma^\mathrm{\parallel}_\xi$ &$(1/T_1)_{\mathrm{exp}, \xi}$ & $(1/T_1)_{\mathrm{cal}, \xi}$  \\
&  ($10^{12} s^{-2}$) & ($10^{12} s^{-2}$) &  (s$^{-1}$) & (s$^{-1}$) \\ \hline
4 T & & & & \\ \hline 
$a$ & $\mathbf{7074}$ & 0 & 10680 & - \\
$b$ & $\mathbf{3263}$ & 1126 & 2039 & 5294 \\
$c$ & $\mathbf{3959}$ & 1372 & 3317 & 6425 \\ \hline
10 T & & & & \\ \hline 
$a$ & 7187 & $\mathbf{0}$ & 264 & - \\
$b$ & 3503 & $\mathbf{1025}$ & 527 & 544.2 \\
$c$ & 4327 & $\mathbf{1216}$ & 524 & 652.1 \\
\hline
\end{tabular}
\end{center}
\end{table}

\subsection{Field dependence of anisotropic spin fluctuations}
Based on the symmetry properties of hyperfine interaction at $^{7}$Li and $^{51}$V nuclei we have succeeded 
to separately determine the transverse and longitudinal spin fluctuations, 
$\langle S_\perp (\mathbf{q}, \omega) \rangle$ and $\langle S_\parallel (\mathbf{q}, \omega) \rangle$, 
using the experimental data of $1/T_1$ as shown in Fig.~\ref{SF}. At 4~T 
$\langle S_\perp (\mathbf{q}, \omega) \rangle$ is much larger than 
$\langle S_\parallel (\mathbf{q}, \omega) \rangle$, while opposite result 
$\langle S_\parallel (\mathbf{q}, \omega) \rangle \gg \langle S_\perp (\mathbf{q}, \omega) \rangle$
is obtained at 10~T. This change of the behavior is naturally understood if one consider the magnetic ordering
in the ground states. The helical and SDW states appear at 4~T and 10~T, respectively, and fluctuations 
of the order parameter are expected to be dominant near $T_N$, consistent with the observation.

What is remarkable though is the pronounced asymmetry in the field dependence of  
$\langle S_\perp (\mathbf{q}, \omega) \rangle$ and $\langle S_\parallel (\mathbf{q}, \omega) \rangle$. 
The longitudinal fluctuation $\langle S_\parallel (\mathbf{q}, \omega) \rangle$ is enhanced only slightly
when the ground state changes from helical to SDW states. On the other hand, the transverse 
fluctuation $\langle S_\perp (\mathbf{q}, \omega) \rangle$ is suppressed very strongly (by a factor of 
40) when the field is increased form 4 to 10~T.  Thus we conclude that the transition from the helical
to the SDW phase is driven by the suppression of the transverse spin correlation, not by the enhancement 
of the longitudinal correlation. 

This conclusion is further reinforced by examining the temperature dependence of 
$\langle S_\parallel (\mathbf{q}, \omega) \rangle$ and $\langle S_\perp (\mathbf{q}, \omega) \rangle$, 
which are best represented by $1/T_1$ at $^{7}$Li sites for $H \parallel b$ (Fig.~\ref{T1}b) and $1/T_1$ at $^{51}$V sites for $H \parallel a$ (Fig.~\ref{T1}d), respectively. Figure~\ref{T1}(b) shows that 
the magnitude and the temperature dependence of $\langle S_\parallel (\mathbf{q}, \omega) \rangle$ 
in the paramagnetic state is nearly unchanged at 4 and 10~T. It show a clear peak near $T_N$ at both fields,
indicating growth of longitudinal spin correlation at low temperatures. This correlation does not lead
to a long range SDW order at 4~T because it is overcome by stronger transverse correlation, which 
drives the helical order. On the other hand, $\langle S_\perp (\mathbf{q}, \omega) \rangle$ 
shown in Fig.~\ref{T1}(d) shows remarkably contrasting behavior at 4 and 10~T. While it shows a strong 
divergence towards $T_N$ consistent with the long range helical order,
it is strongly suppressed at low temperatures at 10~T with an activated temperature dependence $1/T_1 \propto \exp(-\Delta/k_B T)$ with the gap $\Delta$ = 1.8~K. 
Such behavior is consistent with theoretical studies on the quasi 1D frustrated chains~\cite{1Dtheory0, 1Dtheory2, 1Dtheory3}.
The theories show that the transverse fluctuations
are the dominant low energy excitations at low fields, where the long-range order of the vector-chirality 
occurs. At high fields, on the other hand, they are expected to develop an energy gap due to formation of 
the bound magnon pairs. This leads to the activated temperature dependence of $1/T_1$
as predicted by Sato \textit{et al.} \cite{1DtheoryofT12}. 
Our results demonstrate that such behavior predicted by the theories are actually observed in $\mathrm{LiCuVO_4}$. 

By using Eq.~\eqref{LiT1uni} and the estimated values of $\langle S_\perp (\mathbf{q}, \omega) \rangle$ and
$\langle S_\parallel (\mathbf{q}, \omega) \rangle$,
we can calculate $1/T_1$ at 4 and 10~T for all field directions. The calculated values are listed in 
Table \ref{coeff} as $(1/T_1)_{\mathrm{cal}}$ and compared with the experimental results $(1/T_1)_{\mathrm{exp}}$. 
The agreement is excellent for the data at 10~T. Although quantitative agreement is not as good at 4~T, the
calculated values $(1/T_1)_{\mathrm{cal}}$ capture correctly the the anisotropy of $(1/T_1)_{\mathrm{exp}}$.  
We think the results are satisfactory, considering the crudeness of the assumption involved. In particular, 
the crystalline anisotropy, i.e. dependence of $\langle S_\perp (\mathbf{q}, \omega) \rangle$ and $\langle S_\parallel (\mathbf{q}, \omega) \rangle$ on the direction of magnetic field, may not be negligible
at low fields near the spin flop transition and could be significant near $T_N$, where 
$\langle S_\perp(\mathbf{q}, \omega) \rangle$ and $\langle S_\parallel(\mathbf{q}, \omega) \rangle$
change very steeply with temperature.

\section{Conclusion}
We have measured NMR spectra and $1/T_1$ at the $^{7}$Li and $^{51}$V nuclei in the quasi 1D frustrated 
magnet $\mathrm{LiCuVO_4}$.
Analysis of NMR spectra based on the symmetry properties of the hyperfine interactions leads us to conclude
that a helically ordered phase appears in the magnetic field of 4~T
and that a SDW phase with longitudinal modulation appears in the magnetic field of 10~T. 
Results of $1/T_1$ in the paramagnetic phase near $T_N$ indicate that dominant low energy spin excitations 
change with magnetic field. The transverse fluctuations are dominant at 4~T corresponding to the helical
order in the ground state. However, they are suppressed at 10~T due to development of an energy gap 
and overcome by the longitudinal fluctuations that lead to the SDW order. These results are consistent 
with the theoretical prediction for the 1D frustrated chains. 

\acknowledgments
We thank C. Michioka, H. Ueda, M. Sato, T. Hikihara, T. Momoi, A. Smerald and N. Shannon for fruitful discussions.
This work was supported by JSPS KAKENHI (B) (No. 21340093, 22350029), the MEXT-GCOE program, 
a Grant-in-Aid for Science Research from Graduate School of Science, Kyoto University,
and carried out under the Visiting Researcher's Program of the Institute for Solid State Physics, The University of Tokyo.

\end{document}